\documentclass[twocolumn,english,showpacs,preprintnumbers,prb]{revtex4-1}
\usepackage{ae,aecompl}
\usepackage[latin9]{inputenc}
\usepackage{amsmath}
\usepackage{amssymb}
\usepackage{graphicx}
\usepackage{esint}

\makeatletter
\@ifundefined{textcolor}{}
{%
 \definecolor{BLACK}{gray}{0}
 \definecolor{WHITE}{gray}{1}
 \definecolor{RED}{rgb}{1,0,0}
 \definecolor{GREEN}{rgb}{0,1,0}
 \definecolor{BLUE}{rgb}{0,0,1}
 \definecolor{CYAN}{cmyk}{1,0,0,0}
 \definecolor{MAGENTA}{cmyk}{0,1,0,0}
 \definecolor{YELLOW}{cmyk}{0,0,1,0}
}

\usepackage{aecompl}
\usepackage{color}

\@ifundefined{textcolor}{}{%
 \definecolor{BLACK}{gray}{0}
 \definecolor{WHITE}{gray}{1}
 \definecolor{RED}{rgb}{1,0,0}
 \definecolor{GREEN}{rgb}{0,1,0}
 \definecolor{BLUE}{rgb}{0,0,1}
 \definecolor{CYAN}{cmyk}{1,0,0,0}
 \definecolor{MAGENTA}{cmyk}{0,1,0,0}
 \definecolor{YELLOW}{cmyk}{0,0,1,0}
}

\usepackage{aecompl}

\@ifundefined{textcolor}{}{%
 \definecolor{BLACK}{gray}{0}
 \definecolor{WHITE}{gray}{1}
 \definecolor{RED}{rgb}{1,0,0}
 \definecolor{GREEN}{rgb}{0,1,0}
 \definecolor{BLUE}{rgb}{0,0,1}
 \definecolor{CYAN}{cmyk}{1,0,0,0}
 \definecolor{MAGENTA}{cmyk}{0,1,0,0}
 \definecolor{YELLOW}{cmyk}{0,0,1,0}
}

\usepackage[active]{srcltx}\usepackage{aecompl}

\@ifundefined{textcolor}{}{%
 \definecolor{BLACK}{gray}{0}
 \definecolor{WHITE}{gray}{1}
 \definecolor{RED}{rgb}{1,0,0}
 \definecolor{GREEN}{rgb}{0,1,0}
 \definecolor{BLUE}{rgb}{0,0,1}
 \definecolor{CYAN}{cmyk}{1,0,0,0}
 \definecolor{MAGENTA}{cmyk}{0,1,0,0}
 \definecolor{YELLOW}{cmyk}{0,0,1,0}
 }

\usepackage{epsfig}\usepackage{dcolumn}\usepackage{bm}

\usepackage{babel}

\makeatother

\usepackage{babel}
\begin{document}

\title{The graphene sheet versus the 2DEG: a relativistic Fano spin-filter
via STM and AFM tips}

\author{A. C. Seridonio$^{1,2}$, E. C. Siqueira$^{2}$, F. M. Souza$^{3}$,
R. S. Machado$^{2}$, S. S. Lyra$^{2}$, and I. A. Shelykh$^{4,5}$}

\affiliation{$^{1}$Instituto de Geociências e Ciências Exatas - IGCE, Universidade
Estadual Paulista, Departamento de F\'{i}sica, 13506-970, Rio Claro,
SP, Brazil\\
 $^{2}$Departamento de F\'{i}sica e Qu\'{i}mica, Universidade Estadual
Paulista, 15385-000, Ilha Solteira, SP, Brazil\\
 $^{3}$Instituto de F\'{i}sica, Universidade Federal de Uberlândia,
38400-902, Uberlândia, MG, Brazil\\
 $^{4}$Division of Physics and Applied Physics, Nanyang Technological
University 637371, Singapore\\
 $^{5}$Science Institute, University of Iceland, Dunhagi-3, IS-107,
Reykjavik, Iceland }
\begin{abstract}
We explore theoretically the density of states (LDOS) probed by an
STM tip of 2D systems hosting an adatom and a subsurface impurity,
both capacitively coupled to AFM tips and traversed by antiparallel
magnetic fields. Two kinds of setups are analyzed, a monolayer of
graphene and a two-dimensional electron gas (2DEG). The AFM tips set
the impurity levels at the Fermi energy, where two contrasting behaviors
emerge: the Fano factor for the graphene diverges, while in the 2DEG
it approaches zero. As result, the spin-degeneracy of the LDOS is
lifted exclusively in the graphene system, in particular for the asymmetric
regime of Fano interference. The aftermath of this limit is a counterintuitive
phenomenon, which consists of a dominant Fano factor due to the subsurface
impurity even with a stronger STM-adatom coupling. Thus we find a
full polarized conductance, achievable just by displacing vertically
the position of the STM tip. To the best knowledge, our work is the
first to propose the Fano effect as the mechanism to filter spins
in graphene. This feature arises from the massless Dirac electrons
within the band structure and allows us to employ the graphene host
as a relativistic Fano spin-filter.
\end{abstract}


\maketitle

\section{Introduction}

\label{sec1}

Graphene is a two-dimensional layer of atoms organized in a honeycomb
lattice. Its peculiar band structure consisting of two
Dirac cones placed at the corners of the Brillouin zone and characterized
by a massless relativistic dispersion relation provides the opportunity for scientists to explore relativistic
phenomena in the domain of condensed matter physics. In the past decade graphene
was in focus of the physical community, both theoretical and experimental. In particular,
transport properties of graphene and other carbon- based nanostructures attracted the vivid interest of researchers \cite{key-8,key-9,key-100,key-111,key-11,key-12,key-15,Fazzio,key-444,key-445,key-446,key-447}.

Recent experimental \cite{Eelbo1,Eelbo2} and theoretical \cite{Hardcastle,Virgus,Rudenko,Uchoa,Chan} studies reveal the possibility of the effective absorption of the individual magnetic impurities by single graphene sheets. The presence of such impurities (adatoms) strongly modifies magnetic \cite{Huang,Hong,Lehtinen} and transport properties of graphene \cite{Chen,Pi,Alemani} which can be used for variety of technological applications including chemical sensing \cite{Schedin,Wehling}. The convenient experimental techniques for investigation of the properties of individual adatoms is provided by Scanning Tunneling Microscope (STM) \cite{Gyamfi1,Gyamfi2,Brar}. The latter is made by a metallic tip that probes, for low enough temperatures, the local density of states (LDOS) of a sample by measuring the differential conductance \cite{key-1,key-2}. In this scenario, the STM of impurities adsorbed on graphene reveals the scattering of electrons in a relativistic environment.

\begin{figure}[h]
\includegraphics[width=0.43\textwidth]{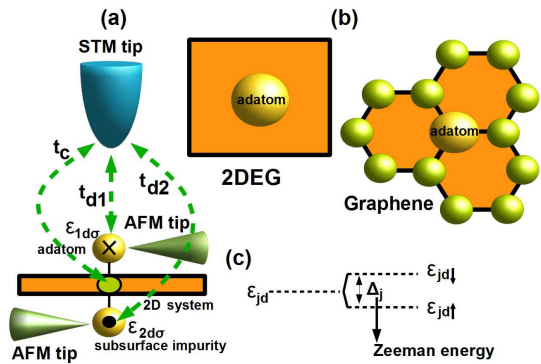} \caption{\label{fig:Pic1}(Color online) (a) It is shown a schematic diagram of the system
considered in this paper. An STM tip is coupled to the atom adsorbed
(adatom) in a 2D host which may be a 2D electron gas (2DEG) or a graphene
sheet. A subsurface impurity is also considered which lies beneath
the 2D system. The AFM tips allow us to control the energy levels
of the impurities which are under the presence of antiparallel magnetic
fields. The cross and dot at each impurity indicate the orientation
of the magnetic field. (b) The top views of the setups considered:
a 2DEG and a graphene sheet, respectively. (c) The Zeeman splitting
caused by the magnetic field in both impurities.}
\end{figure}

The LDOS of metallic systems coupled to impurities displays Fano profiles \cite{Fano1,Fano2}
resulting from the quantum interference between competing channels
in the electron transport. Such an effect arises from the interplay
between the paths of itinerant electrons that travel from the tip directly to the conduction band
of the host and those that tunnel via impurity. The total Fano factor, established by the superposition
of these electron paths, defines the shape of the profiles of the differential conductance.

In the last decade particular attention has been devoted to the Fano
effect in regular metals with magnetic adatoms in the Kondo limit
\cite{Hewson}. In this situation, the description of the host as
a two-dimensional electron gas (2DEG) has been successfully verified
\cite{STM4,STM5,STM6,STM12,STM13}.

Additionally, in the emerging field of spintronics, the presence of spin-polarized hosts gives rise to interesting new features \cite{SPSTM5,Kawahara,Yunong,FM88,Poliana,FM9,FM10,FM11,FM12,new1,new2,new3,new4,new5,SPSTM1,SPSTM2,Seridonio}.
For these cases, the splitting of the Kondo peak in the conductance
characterizes the fingerprint of itinerant magnetism in the host \cite{Seridonio,Kawahara}.
A spin-polarized tip and a nonmagnetic host
also lead to a spin-dependent STM setup. Particulary, the aforementioned
system behaves either as a spin-filter in the Kondo regime or as a
spin-diode away from it \cite{SPSTM1,SPSTM2,Poliana}.
Moreover, a Fano spin-filter can in principle be realized even in the absence of ferromagnetism
and Kondo effect. This can be achieved in the side-coupled geometry
of a quantum wire hybridized with a quantum dot (QD) where spin- degeneracy is lifted by the external magnetic field applied at the QD region \cite{key-10}.

The properties of individual magnetic adatoms hosted by graphene were previously investigated theoretically by using the
single-impurity Anderson Hamiltonian \cite{SIAM}, both  in the regime of high
temperatures $T\gg T_{K}$ (the Kondo temperature) when
Hartree-Fock approach can be used \cite{key-11,key-12} and for $T\ll T_{K}$ when Kondo correlations become important. In the latter case by changing the adatom level in the vicinity of the Fermi energy,
it has been predicted that the Kondo peak arises in a narrower energy
range than in normal metals \cite{key-15}. This is due to the difference of the dispersion of the carriers in the two systems: while the 2DEG is described by a parabolic dispersion, the honeycomb
lattice of graphene leads to a linear dispersion relation near the
Fermi level.

In the current paper we compare further the manifestations of spin- related phenomena in normal metals and graphene in the geometry of the two side- coupled impurities traversed by antiparallel
magnetic fields (Figs. \ref{fig:Pic1}(a) and (b)).
These fields introduce a Zeeman splitting of the impurities levels as depicted in Fig. \ref{fig:Pic1}(c). Additionally,
two atomic force microscope (AFM) tips are capacitively coupled to
the impurities in order to set their energies at the host Fermi energy \cite{Fazzio}.
These tips play the role of metallic gates usually employed to tune
the levels of QDs embedded in nanostructures \cite{QD1,QD2}. Notice that the electrons from the tip
are able to tunnel directly to three different sites with different
amplitudes denoted as: tunneling tip-host ($t_{c}$), tunneling tip-adatom
($t_{d1}$) and tip-subsurface impurity ($t_{d2}$).

The quantum interference between the alternative paths taken by
the electrons rules the transport
through the system and leads to the typical Fano shape of the profiles of the differential conductance of the system. They can be characterized in terms of Fano parameters which allow us to determine the relative impact of each path
into the global response probed by the STM tip. The Fano parameters
are dependent on the properties of the host and demonstrate opposite behavior in graphene and normal metals. In the former case in the vicinity of the host Fermi energy the Fano factor diverges, while in the latter case it approaches to zero. In order to explore such contrasting features, in our further consideration we set the levels of the impurities at the Fermi
energy. As it will be shown in this paper, the lifting of the spin-degeneracy of the LDOS is only feasible
for the graphene system, in particular, in the asymmetric limit of
Fano parameters. In this regime, a counterintuitive phenomenon is revealed,
which is due to the Fano factor of the subsurface impurity that dominates
the interference even with a stronger STM-adatom coupling. We
find that the majority spin component of the LDOS can be tuned by
displacing vertically the STM tip towards (or away from) the host.
We also demonstrate that there is an STM tip position where the conductance becomes full
polarized. The graphene host thus allows us to emulate
an ideal relativistic Fano spin-filter on massless Dirac
fermions.

In order to model the system illustrated in Fig. \ref{fig:Pic1} an
approach based on the two-impurity Anderson Hamiltonian and going beyond Hartree-Fock approximation and valid away from the Kondo regime
was developed. We consider both cases of 2DEG and graphene monolayer. In the latter
system, we follow the approach proposed in Ref.~{[}\onlinecite{key-15}{]},
where an impurity is adsorbed above a single site of the host. Additionally,
we take into account a subsurface impurity, situated opposite to
the adatom. By using equation-of-motion technique for the Green's functions we derive a spin-resolved formula for
the LDOS, characterized by Fano interference parameters.

The paper is organized as follows. In Sec. \ref{sec2} we present the
theoretical model of metallic surfaces with two impurities and derive
the expression for spin-dependent LDOS for the setups shown at Fig. \ref{fig:Pic1}. The
decoupling scheme Hubbard I \cite{book2} for the calculation of the
Green's functions is presented in Sec. \ref{sec3}. In Sec. \ref{sec4}
we derive the expressions of the noninteracting self-energies
of the impurities as well as the Fano parameters for the graphene
sheet and the 2DEG, respectively. The results of the calculations are presented and discussed in Sec. \ref{sec5}.
Conclusions are summarized in Sec. \ref{sec6}.

\section{Theoretical Model}

\label{sec2}

\subsection{HAMILTONIAN }

\label{sub:sec2A}

In order to probe the LDOS of metallic surfaces, we consider an STM
tip weakly connected to hosts hybridized with a pair of side-coupled
impurities as outlined in Fig. \ref{fig:Pic1}. The systems we investigate
are described according to the Anderson model \cite{key-15} given
by the Hamiltonian
\begin{align}
\mathcal{H}^{2D} & =\sum_{s\sigma}\int dk\varepsilon_{k}c_{sk\sigma}^{\dagger}c_{sk\sigma}+\sum_{j\sigma}\varepsilon_{jd\sigma}d_{j\sigma}^{\dagger}d_{j\sigma}\nonumber \\
 & +\sum_{j}\mathcal{U}_{j}n_{j\uparrow}n_{j\downarrow}+\sum_{js\sigma}\int dk\mathcal{V}_{jk}(c_{sk\sigma}^{\dagger}d_{j\sigma}+\text{H.c.}).\label{eq:TIAM}
\end{align}

The surface electrons forming the hosts are described by the operators
$c_{sk\sigma}^{\dagger}$ ($c_{sk\sigma}$) for the creation (annihilation)
of an electron in a quantum state labeled by the wave number $k$,
spin $\sigma$ and in the case of the graphene, an additional index
$s$ standing for the valley index \cite{key-111}. For the 2DEG,
the quantum number $s$ does not exist. The dispersion relation for
the graphene electrons is
\begin{equation}
\varepsilon_{k}=\hbar v_{F}k,\label{eq:disper1}
\end{equation}
with $\hbar$ as the Planck constant divided by $2\pi$ and $v_{F}$
as the Fermi velocity. For the impurities, $d_{j\sigma}^{\dagger}$
($d_{j\sigma}$) creates (annihilates) an electron with spin $\sigma$
in the state $\varepsilon_{jd\sigma}$, with the index $j=1,2$ corresponding to the upper and lower impurities.

The magnetic fields split the energies $\varepsilon_{jd\sigma}$:
\begin{equation}
\varepsilon_{jd\sigma}=\varepsilon_{jd}-\sigma\frac{\Delta_{j}}{2},\label{eq:E2_up_down}
\end{equation}
where $\Delta_{j}$ is the Zeeman energy. Here we employ antiparallel
magnetic fields established by the condition $\Delta_{1}=-\Delta_{2}$.

The third term in Eq.~(\ref{eq:TIAM}) accounts for the on-site Coulomb
interaction $U_{j}$, with $n_{j\sigma}=d_{j\sigma}^{\dagger}d_{j\sigma}$.
Finally, the last two terms mix the host continuum of states of the
conduction band and the levels $\varepsilon_{jd\sigma}$, where $\text{H.c.}$
stands for the Hermitian conjugate of the first term. This hybridization
occurs at the impurity sites via the coupling
\begin{equation}
\mathcal{V}_{jk}=\frac{v_{0}}{2\pi}\sqrt{\frac{\pi\Omega_{0}}{\mathcal{N}}}\sqrt{\left|k\right|},\label{eq:c1}
\end{equation}
where $\mathcal{N}$ is the number of conduction states, the parameters
$v_{0}$ and $\Omega_{0}$ denote the host-impurity hybridization
in energy dimensions and the unit cell area, respectively.

The densities of states of the hosts per spin are different for graphene and 2DEG and are given by the expressions
\begin{equation}
\rho_{0}=\rho^{GS}\left(\varepsilon\right)=\sum_{s}\frac{\Omega_{0}}{2\pi}\frac{\left|\varepsilon\right|}{\left(\hbar v_{F}\right)^{2}}=\frac{\left|\varepsilon\right|}{D^{2}}\label{eq:LDOS_free}
\end{equation}
and

\begin{equation}
\rho_{0}=\rho^{2DEG}\left(\varepsilon\right)=\frac{\Omega_{0}}{2\pi}\frac{D}{\left(\hbar v_{F}\right)^{2}}=\frac{1}{2D},\label{eq:LDOS_free_2}
\end{equation}
where $D$ denotes the band-edge.

\subsection{LDOS for the ``host+impurities'' system in presence of STM and
AFM tips}

\label{sub:sec2B}

By applying the linear response theory, in which the
STM tip is considered as a probe, it is possible to show that the differential conductance
per spin is determined by \cite{key-12}
\begin{equation}
\mathcal{G}^{\sigma}(V)\sim\frac{e^{2}}{h}\pi\Gamma_{c}\rho_{LDOS}^{\sigma}\left(eV\right),\label{eq:DC}
\end{equation}
where $e$ is the electron charge, $\Gamma_{c}=4\pi t_{c}^{2}\rho_{tip}$,
$t_{c}$ is the tunneling term between the STM tip and the host, $\rho_{tip}$
is the DOS for the tip, $V$ is the bias-voltage and $\rho_{LDOS}^{\sigma}$
is spin-resolved LDOS of the ``host+impurities'' system.

To obtain the LDOS, we introduce the retarded Green's function
\begin{align}
\mathcal{R}_{\sigma}\left(t\right) & =-\frac{i}{\hbar}\theta\left(t\right){\tt Tr}\{\varrho_{2D}\left[\psi_{\sigma}\left(t\right),\psi_{\sigma}^{\dagger}\left(0\right)\right]_{+}\}\label{eq:PSI_R}
\end{align}
in the time domain, where $\theta\left(t\right)$ is the Heaviside
function, $\varrho_{2D}$ is the density matrix
of the system described by the Hamiltonian {[}Eq. (\ref{eq:TIAM}){]} and $[\cdots,\cdots]_{+}$
is the anticommutator of the field operator taken in Heisenberg picture \cite{Seridonio}
\begin{multline}
\psi_{\sigma}\left(t\right)=\frac{1}{2\pi}\sqrt{\frac{\pi\Omega_{0}}{\mathcal{N}}}\sum_{s}\int\sqrt{\left|k\right|}dkc_{sk\sigma}(t)\\+\left(\pi\rho_{0}v_{0}\right)\sum_{j}q_{j}^{e}d_{j}(t),\label{eq:PSI_R-1-1}
\end{multline}
with
\begin{equation}
q_{j}^{e}=\left(\pi\rho_{0}v_{0}\right)^{-1}\left(\frac{t_{dj}}{t_{c}}\right)\label{eq:Fano2}
\end{equation}
being the extrinsic Fano factor, defined by the couplings between the STM
tip and the ``host+impurities'' system.

From Eq.~(\ref{eq:PSI_R}), the spin-resolved LDOS of the host can
be obtained as
\begin{equation}
\rho_{LDOS}^{\sigma}=-\frac{1}{\pi}{\tt Im}(\tilde{\mathcal{R}}_{\sigma}),\label{eq:FM_LDOS}
\end{equation}
where $\tilde{\mathcal{R}}_{\sigma}$ is the Fourier transform
of $\mathcal{R}_{\sigma}(t)$. To obtain an analytical expression
for the LDOS, we apply the equation-of-motion approach to the Eq.
(\ref{eq:PSI_R}). Substituting Eq. (\ref{eq:PSI_R-1-1})
in Eq. (\ref{eq:PSI_R}), one gets
\begin{align}
\mathcal{R}_{\sigma}(t) & =\left(\frac{1}{2\pi}\sqrt{\frac{\pi\Omega_{0}}{\mathcal{N}}}\right)^{2}\sum_{s\tilde{s}}\int\sqrt{\left|k\right|}dk\sqrt{\left|q\right|}dq\mathcal{\mathcal{R}}_{c_{sk}c_{\tilde{s}q}}^{\sigma}\nonumber \\
 & +\left(\pi\rho_{0}v_{0}\right)\sum_{js}q_{j}^{e}\left(\frac{1}{2\pi}\sqrt{\frac{\pi\Omega_{0}}{\mathcal{N}}}\right)\int\sqrt{\left|k\right|}dk\nonumber \\
 & \times(\mathcal{R}{}_{d_{j}c_{sk}}^{\sigma}+\mathcal{R}_{c_{sk}d_{j}}^{\sigma})+\left(\pi\rho_{0}v_{0}\right)^{2}\sum_{jl}q_{j}^{e}q_{l}^{e}\mathcal{R}_{d_{j}d_{l}}^{\sigma},\nonumber \\
\label{eq:GF_1}
\end{align}
expressed in terms of the Green's functions $\mathcal{\mathcal{R}}_{c_{sk}c_{\tilde{s}q}}^{\sigma}$,
$\mathcal{R}_{d_{j}c_{sk}}^{\sigma}$, $\mathcal{R}_{c_{sk}d_{j}}^{\sigma}$
and $\mathcal{R}_{d_{j}d_{l}}^{\sigma}$.

First, we have to determine
\begin{align}
\mathcal{R}_{c_{sk}c_{\tilde{s}q}}^{\sigma}\left(t\right) & =-\frac{i}{\hbar}\theta\left(t\right){\tt Tr}\{\varrho_{2D}[c_{sk\sigma}\left(t\right),c_{\tilde{s}q\sigma}^{\dagger}\left(0\right)]_{+}\}\label{eq:GF_2}
\end{align}
by acting by the operator~$\partial_{t}\equiv\frac{\partial}{\partial t}$ on Eq.
(\ref{eq:GF_2}). We find
\begin{eqnarray}
\partial_{t}\mathcal{R}_{c_{sk}c_{\tilde{s}q}}^{\sigma}\left(t\right) & = & -\frac{i}{\hbar}\delta\left(t\right){\tt Tr}\{\varrho_{2D}[c_{sk\sigma}\left(t\right),c_{\tilde{s}q\sigma}^{\dagger}\left(0\right)]_{+}\}\nonumber \\
 & - & \frac{i}{\hbar}\varepsilon_{k}\mathcal{R}_{c_{sk}c_{\tilde{s}q}}^{\sigma}\left(t\right)-\frac{i}{\hbar}\sum_{j}\mathcal{V}_{jk}\mathcal{R}_{d_{j}c_{\tilde{s}q}}^{\sigma}\left(t\right),\nonumber \\
\label{eq:GF_3}
\end{eqnarray}
where we have used
\begin{align}
i\hbar\partial_{t}c_{sk\sigma}\left(t\right) & =[c_{sk\sigma},\mathcal{H}^{2D}]=\varepsilon_{k}c_{sk\sigma}\left(t\right)+\sum_{j}\mathcal{V}_{jk}d_{j\sigma}\left(t\right).\label{eq:HB_I}
\end{align}

In the energy domain, we solve Eq. (\ref{eq:GF_3}) for $\tilde{\mathcal{R}}_{c_{sk}c_{\tilde{s}q}}^{\sigma}$
and obtain

\begin{align}
\tilde{\mathcal{R}}_{c_{sk}c_{\tilde{s}q}}^{\sigma} & =\frac{\delta\left(k-q\right)\delta_{s\tilde{s}}}{\varepsilon^{+}-\varepsilon_{k}}+\sum_{j}\frac{\mathcal{V}_{jk}}{\varepsilon^{+}-\varepsilon_{k}}\tilde{\mathcal{R}}_{d_{j}c_{sq}}^{\sigma},\label{eq:GF_4}
\end{align}
where $\varepsilon^{+}=\varepsilon+i\eta$ and $\eta\rightarrow0^{+}$.
Notice that we also need to calculate the mixed Green's function $\tilde{\mathcal{R}}_{d_{j}c_{sq}}^{\sigma}.$
To this end, we define the advanced Green's function
\begin{align}
\mathcal{A}_{d_{j}c_{sq}}^{\sigma}\left(t\right) & =\frac{i}{\hbar}\theta\left(-t\right){\tt Tr}\{\varrho_{2D}[d_{j\sigma}^{\dagger}\left(0\right),c_{sq\sigma}\left(t\right)]_{+}\},\label{eq:GF_5}
\end{align}
whose equation of motion reads,
\begin{align}
\partial_{t}\mathcal{A}_{d_{j}c_{sq}}^{\sigma}\left(t\right) & =-\frac{i}{\hbar}\delta\left(t\right){\tt Tr}\{\varrho_{2D}[d_{j\sigma}^{\dagger}\left(0\right),c_{sq\sigma}\left(t\right)]_{+}\}\nonumber \\
 & -\frac{i}{\hbar}\varepsilon_{q}\mathcal{A}_{d_{j}c_{sq}}^{\sigma}\left(t\right)-\frac{i}{\hbar}\sum_{l}\mathcal{V}_{lq}\mathcal{A}_{d_{j}d_{l}}^{\sigma}\left(t\right),\label{eq:GF_6}
\end{align}
where we have used once again Eq. (\ref{eq:HB_I}), interchanging
$k\leftrightarrow q$. The Fourier transform of Eq.~(\ref{eq:GF_6})
leads to
\begin{align}
\varepsilon^{-}\tilde{\mathcal{A}}_{d_{j}c_{sq}}^{\sigma} & =\varepsilon_{q}\tilde{\mathcal{A}}_{d_{j}c_{sq}}^{\sigma}+\sum_{l}\mathcal{V}_{lq}\tilde{\mathcal{A}}_{d_{j}d_{l}}^{\sigma},\label{eq:GF_7}
\end{align}
with $\varepsilon^{-}=\varepsilon-i\eta$. Applying the property $\tilde{\mathcal{R}}_{d_{j}c_{sq}}^{\sigma}=(\tilde{\mathcal{A}}_{d_{j}c_{sq}}^{\sigma})^{\dagger}$
on Eq. (\ref{eq:GF_7}), we show that
\begin{align}
\varepsilon^{+}\tilde{\mathcal{R}}_{d_{j}c_{sq}}^{\sigma} & =\varepsilon_{q}\tilde{\mathcal{R}}_{d_{j}c_{sq}}^{\sigma}+\sum_{l}\mathcal{V}_{lq}\tilde{\mathcal{R}}_{d_{j}d_{l}}^{\sigma},\label{eq:GF_8}
\end{align}

\begin{equation}
\tilde{\mathcal{R}}_{d_{j}c_{sq}}^{\sigma}=\sum_{l}\frac{\mathcal{V}_{lq}}{\varepsilon^{+}-\varepsilon_{q}}\tilde{\mathcal{R}}_{d_{j}d_{l}}^{\sigma}\label{eq:GF_9}
\end{equation}
and analogously,
\begin{equation}
\tilde{\mathcal{R}}_{c_{sq}d_{j}}^{\sigma}=\sum_{l}\frac{\mathcal{V}_{lq}}{\varepsilon^{+}-\varepsilon_{q}}\tilde{\mathcal{R}}_{d_{l}d_{j}}^{\sigma}.\label{eq:GF_10}
\end{equation}

Now we substitute Eq. (\ref{eq:GF_9}) into Eq. (\ref{eq:GF_4}) and the latter, together with Eq. (\ref{eq:GF_10}), into Eq. (\ref{eq:GF_1}) and
determine
\begin{align}
\tilde{\mathcal{R}}_{\sigma} & =\left(\frac{1}{2\pi}\sqrt{\frac{\pi\Omega_{0}}{\mathcal{N}}}\right)^{2}\sum_{s}\int kdk\frac{1}{\varepsilon^{+}-\varepsilon_{k}}\nonumber \\
 & +\left(\pi\rho_{0}v_{0}\right)^{2}\sum_{jl}(q_{j}-i\mathcal{F}_{j})\tilde{\mathcal{R}}_{d_{j}d_{l}}^{\sigma}(q_{l}-i\mathcal{F}_{l})\nonumber \\
 & +\left(\pi\rho_{0}v_{0}\right)^{2}\sum_{jl}q_{j}^{e}(q_{l}-i\mathcal{F}_{l})(\tilde{\mathcal{R}}_{d_{j}d_{l}}^{\sigma}+\tilde{\mathcal{R}}_{d_{l}d_{j}}^{\sigma})\nonumber \\
 & +\left(\pi\rho_{0}v_{0}\right)^{2}\sum_{jl}q_{j}^{e}q_{l}^{e}\tilde{\mathcal{R}}_{d_{j}d_{l}}^{\sigma},\label{eq:GFN}
\end{align}
where
\begin{equation}
q_{j}=\frac{1}{\pi\rho_{0}v_{0}^{2}}{\tt Re}\Sigma_{jj}\label{eq:Fano_j}
\end{equation}
is the Fano parameter \cite{key-12} due to the host-impurity coupling
and
\begin{align}
\mathcal{F}_{j} & =-\frac{1}{\pi\rho_{0}v_{0}^{2}}{\tt Im}\Sigma_{jj},\label{eq:Friedel_j}
\end{align}
with
\begin{equation}
\Sigma{}_{l\tilde{l}}=\sum_{s}\int dk\frac{\mathcal{V}_{lk}\mathcal{V}_{\tilde{l}k}}{\varepsilon^{+}-\varepsilon_{k}}\label{eq:r_s_e}
\end{equation}
being the noninteracting self-energy of the impurities \cite{key-12}.
From Eqs.~(\ref{eq:FM_LDOS}) and (\ref{eq:GFN}), we finally derive
the spin-resolved LDOS
\begin{equation}
\rho_{LDOS}^{\sigma}=\rho_{1122}^{\sigma}+\rho_{1221}^{\sigma},\label{eq:LDOS_split}
\end{equation}
where
\begin{align}
\rho_{1122}^{\sigma} & =\rho_{0}+\rho_{0}\Gamma\sum_{j}[(\mathcal{F}_{j}^{2}-q_{Tj}^{2}){\tt Im}(\tilde{\mathcal{R}}_{d_{j}d_{j}}^{\sigma})\nonumber \\
 & +2q_{Tj}\mathcal{F}_{j}{\tt Re}(\tilde{\mathcal{R}}_{d_{j}d_{j}}^{\sigma})]\label{eq:LDOS_p1}
\end{align}
is the \emph{direct term} of the LDOS and
\begin{align}
\rho_{1221}^{\sigma} & =\rho_{0}\Gamma\sum_{j\neq l}[(\mathcal{F}_{j}\mathcal{F}_{l}-q_{Tj}q_{Tl}){\tt Im}(\tilde{\mathcal{R}}_{d_{j}d_{l}}^{\sigma})\nonumber \\
 & +(q_{Tj}\mathcal{F}_{l}+q_{Tl}\mathcal{F}_{j}){\tt Re}(\tilde{\mathcal{R}}_{d_{j}d_{l}}^{\sigma})\label{eq:LDOS_p2}
\end{align}
represents the \emph{mixing term} that arises from the interference
between the impurities, with
\begin{equation}
q_{Tj}=q_{j}+q_{j}^{e}\label{eq:Fano_Total}
\end{equation}
being the total Fano factor and $\Gamma=\pi v_{0}^{2}\rho_{0}$ is the
Anderson parameter.

Eq.~(\ref{eq:LDOS_split}) is the main analytical result of this
paper. It describes the spin-resolved LDOS of 2D systems with two
impurities in the side-coupled geometry shown at Fig. \ref{fig:Pic1}. This
equation shows the dependence of the LDOS on the direct and mixed Green's functions of the
impurities, $\tilde{\mathcal{R}}_{d_{j}d_{j}}^{\sigma}$ and $\tilde{\mathcal{R}}_{d_{j}d_{l}}^{\sigma}$
respectively, and on the total Fano parameter given by the Eq. (\ref{eq:Fano_Total}).
We highlight that the Zeeman energy of the impurities determines the
spin-dependence of the LDOS and therefore, the spin-filter behavior,
in particular only for the graphene system as we will see.

In order to investigate the spin dependence of the LDOS as well
as the spin-filter effect, we introduce the expression

\begin{equation}
LDOS=\frac{\rho_{LDOS}^{\uparrow}+\rho_{LDOS}^{\downarrow}}{\rho^{GS}\left(D\right)},\label{eq:LDOS_sd}
\end{equation}
for the dimensionless LDOS, where we have used Eq. (\ref{eq:LDOS_free})
for $\rho^{GS}\left(D\right)$ and

\begin{equation}
SP=\frac{\mathcal{G}^{\uparrow}-\mathcal{G}^{\downarrow}}{\mathcal{G}^{\uparrow}+\mathcal{G}^{\downarrow}}\label{eq:SP}
\end{equation}
for the transport polarization of the system settled from Eq. (\ref{eq:DC}).

\section{Green's functions of the impurities}

\label{sec3}

In the present section we calculate $\tilde{\mathcal{R}}_{d_{j}d_{l}}^{\sigma}$
($j,l=1,2$) within the Hubbard I approximation \cite{book2}. This
approach provides reliable results away from the Kondo regime. Thus
we begin by applying the equation-of-motion method on these Green's
functions, which results in

\begin{align}
\left(\varepsilon^{+}-\varepsilon_{ld\sigma}\right)\tilde{\mathcal{R}}_{d_{l}d_{j}}^{\sigma} & =\delta_{lj}+\sum_{\tilde{l}}\Sigma{}_{l\tilde{l}}\tilde{\mathcal{R}}_{d_{\tilde{l}}d_{j}}^{\sigma}\nonumber \\
 & +\mathcal{U}_{l}\tilde{\mathcal{R}}_{d_{l\sigma}n_{d_{l}\bar{\sigma}},d_{j\sigma}}.\label{eq:s1}
\end{align}

In the equation above, $\tilde{\mathcal{R}}_{d_{l\sigma}n_{d_{l}\bar{\sigma}},d_{j\sigma}}$
is a two particle Green's function composed by four fermionic operators,
obtained from the time Fourier transform of
\begin{align}
\mathcal{R}_{d_{l\sigma}n_{d_{l}\bar{\sigma}},d_{j\sigma}} & =-\frac{i}{\hbar}\theta\left(t\right){\tt Tr}\{\varrho_{2D}[d_{l\sigma}\left(t\right)n_{d_{l}\bar{\sigma}}\left(t\right),d_{j\sigma}^{\dagger}\left(0\right)]_{+}\},\label{eq:H_GF}
\end{align}
with $n_{d_{l}\bar{\sigma}}=d_{l\bar{\sigma}}^{\dagger}d_{l\bar{\sigma}}$
and spin $\bar{\sigma}$ (opposite to $\sigma$).

In order to close the system of Green's functions in Eq. (\ref{eq:s1}),
we calculate the time derivative of Eq. (\ref{eq:H_GF}) and then
its time Fourier transform, which leads to

\begin{align}
\left(\varepsilon^{+}-\varepsilon_{ld\sigma}-\mathcal{U}_{l}\right)\tilde{\mathcal{R}}_{d_{l\sigma}n_{d_{l}\bar{\sigma}},d_{j\sigma}} & =\delta_{lj}\left\langle n_{d_{l}\bar{\sigma}}\right\rangle \nonumber \\
+\sum_{s}\int dk\mathcal{V}_{lk} & (\tilde{\mathcal{R}}_{c_{sk\sigma}d_{l\bar{\sigma}}^{\dagger}d_{l\bar{\sigma}},d_{j\sigma}}\nonumber \\
-\tilde{\mathcal{R}}_{c_{sk\bar{\sigma}}^{\dagger}d_{l\bar{\sigma}}d_{l\sigma},d_{j\sigma}} & +\tilde{\mathcal{R}}_{d_{l\bar{\sigma}}^{\dagger}c_{sk\bar{\sigma}}d_{l\sigma},d_{j\sigma}}),\label{eq:H_GF_2}
\end{align}
expressed in terms of new Green's functions of the same order of $\tilde{\mathcal{R}}_{d_{l\sigma}n_{d_{l}\bar{\sigma}},d_{j\sigma}}$
and the occupation number
\begin{equation}
\left\langle n_{d_{l}\bar{\sigma}}\right\rangle =-\frac{1}{\pi}\int_{-D}^{\epsilon_{F}=0}{\tt Im}(\tilde{\mathcal{R}}_{d_{l}d_{l}}^{\bar{\sigma}})d\varepsilon\label{eq:nb}
\end{equation}
determined in accordance with Refs. ~{[}\onlinecite{key-11,key-12}{]}.
By employing the Hubbard I approximation, we decouple the Green's
functions in the right-hand side of Eq. (\ref{eq:H_GF_2}) as follows:
$\tilde{\mathcal{R}}_{c_{sk\bar{\sigma}}^{\dagger}d_{l\bar{\sigma}}d_{l\sigma},d_{j\sigma}}\simeq\left\langle c_{sk\bar{\sigma}}^{\dagger}d_{l\bar{\sigma}}\right\rangle \tilde{\mathcal{R}}_{d_{l}d_{j}}^{\sigma}$
and $\tilde{\mathcal{R}}_{d_{l\bar{\sigma}}^{\dagger}c_{sk\bar{\sigma}}d_{l\sigma},d_{j\sigma}}\simeq\left\langle c_{sk\bar{\sigma}}^{\dagger}d_{l\bar{\sigma}}\right\rangle \tilde{\mathcal{R}}_{d_{l}d_{j}}^{\sigma}$.
As a result, we find

\begin{eqnarray}
\left(\varepsilon^{+}-\varepsilon_{ld\sigma}-\mathcal{U}_{l}\right)\tilde{\mathcal{R}}_{d_{l\sigma}n_{d_{l}\bar{\sigma}},d_{j\sigma}} & = & \delta_{lj}\left\langle n_{d_{l}\bar{\sigma}}\right\rangle \nonumber \\
+\left\langle n_{d_{l}\bar{\sigma}}\right\rangle \left(\sum_{s}\int dk\mathcal{V}_{lk}\right) & \times & \tilde{\mathcal{R}}_{c_{sk\sigma}d_{l\bar{\sigma}}^{\dagger}d_{l\bar{\sigma}},d_{j\sigma}}.\nonumber \\
\label{eq:H_GF_3-1}
\end{eqnarray}

To close the calculation, we need to determine $\tilde{\mathcal{R}}_{c_{sk\sigma}d_{l\bar{\sigma}}^{\dagger}d_{l\bar{\sigma}},d_{j\sigma}}$.
Once again, employing the equation-of-motion approach for $\tilde{\mathcal{R}}_{c_{sk\sigma}d_{l\bar{\sigma}}^{\dagger}d_{l\bar{\sigma}},d_{j\sigma}}$,
we find

\begin{align}
\left(\varepsilon^{+}-\varepsilon_{k}\right)\tilde{\mathcal{R}}_{c_{sk\sigma}d_{l\bar{\sigma}}^{\dagger}d_{l\bar{\sigma}},d_{j\sigma}} & =\mathcal{V}_{lk}\tilde{\mathcal{R}}_{d_{l\sigma}n_{d_{l}\bar{\sigma}},d_{j\sigma}}\nonumber \\
+\sum_{\tilde{s}}\int dq\mathcal{V}_{lq}\tilde{\mathcal{R}}_{c_{sk\sigma}d_{l\bar{\sigma}}^{\dagger}c_{\tilde{s}q\bar{\sigma}},d_{j\sigma}} & +\sum_{\tilde{j}\neq l}\mathcal{V}_{\tilde{j}k}\tilde{\mathcal{R}}_{d_{\tilde{j}\sigma}n_{d_{l}\bar{\sigma}},d_{j\sigma}}\nonumber \\
-\sum_{\tilde{s}}\int dq\mathcal{V}_{lq}\tilde{\mathcal{R}}_{c_{\tilde{s}q\bar{\sigma}}^{\dagger}d_{l\bar{\sigma}}c_{sk\sigma},d_{j\sigma}}.\label{eq:H_GF_4}
\end{align}

For the sake of simplicity, we take the limit $\mathcal{U}_{l}\rightarrow\infty$
and continue with the Hubbard I scheme by making $\tilde{\mathcal{R}}_{c_{sk\sigma}d_{l\bar{\sigma}}^{\dagger}c_{\tilde{s}q\bar{\sigma}},d_{j\sigma}}\simeq\left\langle d_{l\bar{\sigma}}^{\dagger}c_{\tilde{s}q\bar{\sigma}}\right\rangle \tilde{\mathcal{R}}_{c_{sk\sigma}d_{j\sigma}},$
$\tilde{\mathcal{R}}_{c_{\tilde{s}q\bar{\sigma}}^{\dagger}d_{l\bar{\sigma}}c_{sk\sigma},d_{j\sigma}}\simeq\left\langle d_{l\bar{\sigma}}^{\dagger}c_{\tilde{s}q\bar{\sigma}}\right\rangle \tilde{\mathcal{R}}_{c_{sk\sigma}d_{j\sigma}}$
and $\tilde{\mathcal{R}}_{d_{\tilde{j}\sigma}n_{d_{l}\bar{\sigma}},d_{j\sigma}}\simeq\left\langle n_{d_{l}\bar{\sigma}}\right\rangle \tilde{\mathcal{R}}_{d_{\tilde{j}}d_{j}}^{\sigma}$
in Eq. (\ref{eq:H_GF_4}), which becomes

\begin{align}
\tilde{\mathcal{R}}_{c_{sk\sigma}d_{l\bar{\sigma}}^{\dagger}d_{l\bar{\sigma}},d_{j\sigma}} & =\frac{\mathcal{V}_{lk}}{\left(\varepsilon^{+}-\varepsilon_{k}\right)}\tilde{\mathcal{R}}_{d_{l\sigma}n_{d_{l}\bar{\sigma}},d_{j\sigma}}\nonumber \\
 & +\frac{\sum_{\tilde{j}\neq l}\mathcal{V}_{\tilde{j}k}}{\left(\varepsilon^{+}-\varepsilon_{k}\right)}\left\langle n_{d_{l}\bar{\sigma}}\right\rangle \tilde{\mathcal{R}}_{d_{\tilde{j}}d_{j}}^{\sigma}.\label{eq:HG_F5}
\end{align}

Thus by solving the system of Green's functions composed by Eqs. (\ref{eq:s1}),
(\ref{eq:H_GF_3-1}), (\ref{eq:H_GF_4}) and (\ref{eq:HG_F5}), we
obtain

\begin{equation}
\tilde{\mathcal{R}}_{d_{1}d_{1}}^{\sigma}=\frac{1-\left\langle n_{d_{1}\bar{\sigma}}\right\rangle }{\varepsilon-\varepsilon_{1d\sigma}-\sum_{11}-\lambda_{12}^{\bar{\sigma}}\frac{\left(\sum_{11}\right)^{2}}{\varepsilon-\varepsilon_{2d\sigma}-\sum_{11}}},\label{eq:pass3-1}
\end{equation}
where $\lambda_{12}^{\bar{\sigma}}=\left(1-\left\langle n_{d_{1}\bar{\sigma}}\right\rangle \right)\left(1-\left\langle n_{d_{2}\bar{\sigma}}\right\rangle \right)$
and

\begin{equation}
\tilde{\mathcal{R}}_{d_{2}d_{1}}^{\sigma}=\left(1-\left\langle n_{d_{2}\bar{\sigma}}\right\rangle \right)\frac{\sum_{21}}{\varepsilon-\varepsilon_{2d\sigma}-\sum_{21}}\tilde{\mathcal{R}}_{d_{1}d_{1}}^{\sigma},\label{eq:pass9-1}
\end{equation}
with $\sum_{jl}$ determined by Eq. (\ref{eq:r_s_e}). We point out
that the Green's functions $\tilde{\mathcal{R}}_{d_{2}d_{2}}^{\sigma}$
and $\tilde{\mathcal{R}}_{d_{1}d_{2}}^{\sigma}$ can be found by swapping
$1\leftrightarrow2$ in Eqs. (\ref{eq:pass3-1}) and (\ref{eq:pass9-1}).
As a result, Eq. (\ref{eq:pass3-1}) and $\tilde{\mathcal{R}}_{d_{2}d_{2}}^{\sigma}$
allow us to introduce

\begin{equation}
DOS_{jj}^{\sigma}=-\frac{1}{\pi\rho^{GS}\left(D\right)}{\tt Im}(\tilde{\mathcal{R}}_{d_{j}d_{j}}^{\sigma})\label{eq:DOSjj}
\end{equation}
as the dimensionless DOS for the impurities, where we have applied
Eq. (\ref{eq:LDOS_free}) at the band-edge $D$.

\section{Noninteracting self-energies and Fano parameters}

\label{sec4}

In this section we present the calculations of the noninteracting
self-energies of Eq. (\ref{eq:r_s_e}) and the Fano parameters within
the Eqs. (\ref{eq:Fano2}), (\ref{eq:Fano_j}) and (\ref{eq:Fano_Total}).
The Eq. (\ref{eq:r_s_e}) allows us to find

\begin{equation}
\Sigma{}_{l\tilde{l}}=\Sigma{}_{l\tilde{l}}^{GS}=\eta^{GS}\left(\varepsilon\ln\left|\frac{\varepsilon^{2}}{\varepsilon^{2}-D^{2}}\right|-i\pi\left|\varepsilon\right|\right)\label{eq:s_e_GS}
\end{equation}
for the graphene sheet, with
\begin{equation}
\eta^{GS}=\frac{\Omega_{0}}{2\pi\mathcal{N}}\frac{v_{0}^{2}}{\left(\hbar v_{F}\right)^{2}}=\frac{v_{0}^{2}}{D^{2}}\label{eq:eta_GS}
\end{equation}
and
\begin{equation}
\Sigma{}_{l\tilde{l}}=\Sigma{}_{l\tilde{l}}^{2DEG}=\eta^{2DEG}\left(D\ln\left|\frac{D+\varepsilon}{D-\varepsilon}\right|-i\pi D\right)\label{eq:s_e_GS-2}
\end{equation}
for the case of the 2DEG, where
\begin{equation}
\eta^{2DEG}=\frac{\Omega_{0}}{2\pi\mathcal{N}}\frac{v_{0}^{2}}{\left(\hbar v_{F}\right)^{2}}=\frac{v_{0}^{2}}{2D^{2}}.\label{eq:eta_2DEG}
\end{equation}
Notice that for $\varepsilon\ll D$, ${\tt Re}(\Sigma{}_{l\tilde{l}}^{GS})={\tt Re}(\Sigma{}_{l\tilde{l}}^{2DEG})\rightarrow0$.

The self-energy of Eq. (\ref{eq:s_e_GS}) is in accordance with the
corresponding determined in Refs. ~{[}\onlinecite{key-11,key-12,key-15}{]}.
For the 2DEG, we recover the result found in Ref. ~{[}\onlinecite{Hewson}{]}.

From Eqs. (\ref{eq:Fano2}), (\ref{eq:Fano_j}), (\ref{eq:Friedel_j})
and (\ref{eq:Fano_Total}), we determine

\begin{equation}
q_{Tj}^{GS}=\frac{1}{\pi}\ln\left|\frac{\varepsilon^{2}}{\varepsilon^{2}-D^{2}}\right|+\left(\frac{t_{dj}}{t_{c}}\right)\frac{D^{2}}{\pi\nu_{0}}\frac{1}{\left|\varepsilon\right|}\label{eq:Fano_jGS}
\end{equation}
and

\begin{equation}
\mathcal{F}_{j}^{GS}=1\label{eq:A_GS-1}
\end{equation}
for the graphene system, while for the 2DEG we have
\begin{equation}
q_{Tj}^{2DEG}=\frac{1}{\pi}\ln\left|\frac{D+\varepsilon}{D-\varepsilon}\right|+\left(\frac{t_{dj}}{t_{c}}\right)\frac{2D}{\pi\nu_{0}}\label{eq:Fano_j2DEG}
\end{equation}
and

\begin{equation}
\mathcal{F}_{j}^{2DEG}=1.\label{eq:A_2DEG}
\end{equation}

We emphasize that for $t_{dj}/t_{c}\ll1$ and energies $\varepsilon\ll D$,
Eqs. (\ref{eq:Fano_jGS}) and (\ref{eq:Fano_j2DEG}) exhibit opposite
behaviors: $\left|q_{Tj}^{GS}\right|\rightarrow\infty$ and $\left|q_{Tj}^{2DEG}\right|\rightarrow0$.

\section{NUMERICAL RESULTS}

\label{sec5}

The present approach is for $T\gg T_{K}$ and within a range of temperatures where we can safely define the Heaviside step
function as the Fermi distribution in the host. This assumption was previously considered in the Refs. ~{[}\onlinecite{key-11,key-12}{]}.
We measure the densities and energies in units of $\rho^{GS}\left(D\right)$
and $\left[\rho^{GS}\left(D\right)\right]^{-1}$, respectively, where
we have used Eq. (\ref{eq:LDOS_free}) at the band-edge $D=7$ eV
\cite{key-11,key-12}. For the Zeeman splittings, we employ $\Delta_{1}=-\Delta_{2}=2\times10^{-5}$
in Eq. (\ref{eq:E2_up_down}) corresponding to a magnetic field of
$\approx2.4$ T. We have also used $v_{0}=0.14$ in Eq. (\ref{eq:c1})
\cite{key-11,key-12}. The levels of the impurities are set to $\varepsilon_{1d}=\varepsilon_{2d}=0$.

\subsection{Graphene system}

\label{sub:5a}

Here we analyze the graphene system by employing Eq. (\ref{eq:LDOS_sd})
in combination with Eqs. (\ref{eq:LDOS_free}), (\ref{eq:pass3-1}),
(\ref{eq:pass9-1}), (\ref{eq:s_e_GS}), (\ref{eq:Fano_jGS}) and (\ref{eq:A_GS-1}).
\begin{figure}[h]
\centerline{\resizebox{3.5in}{!}{ \includegraphics[clip,width=0.6\textwidth]{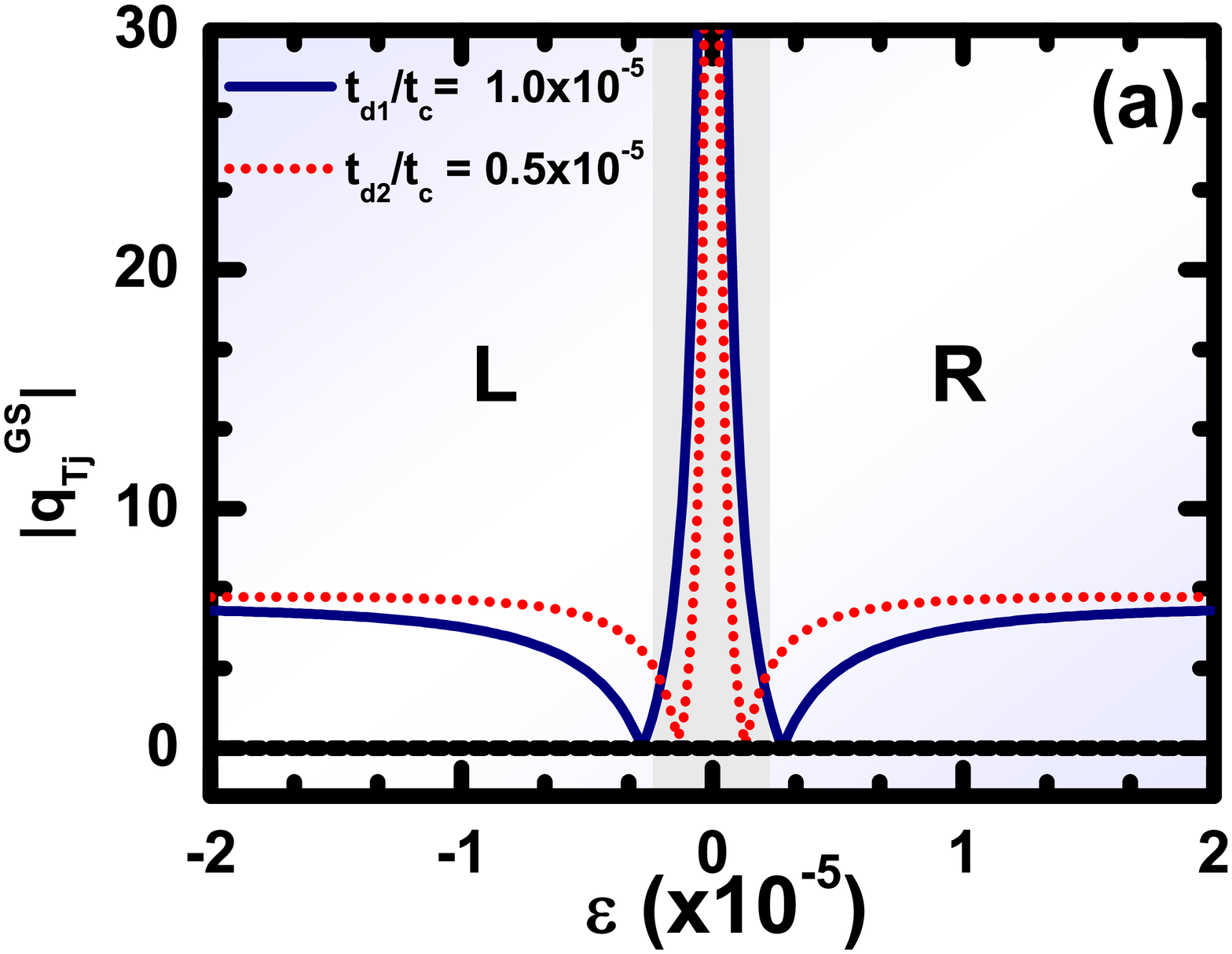}}}
\centerline{\resizebox{3.5in}{!}{ \includegraphics[clip,width=0.6\textwidth]{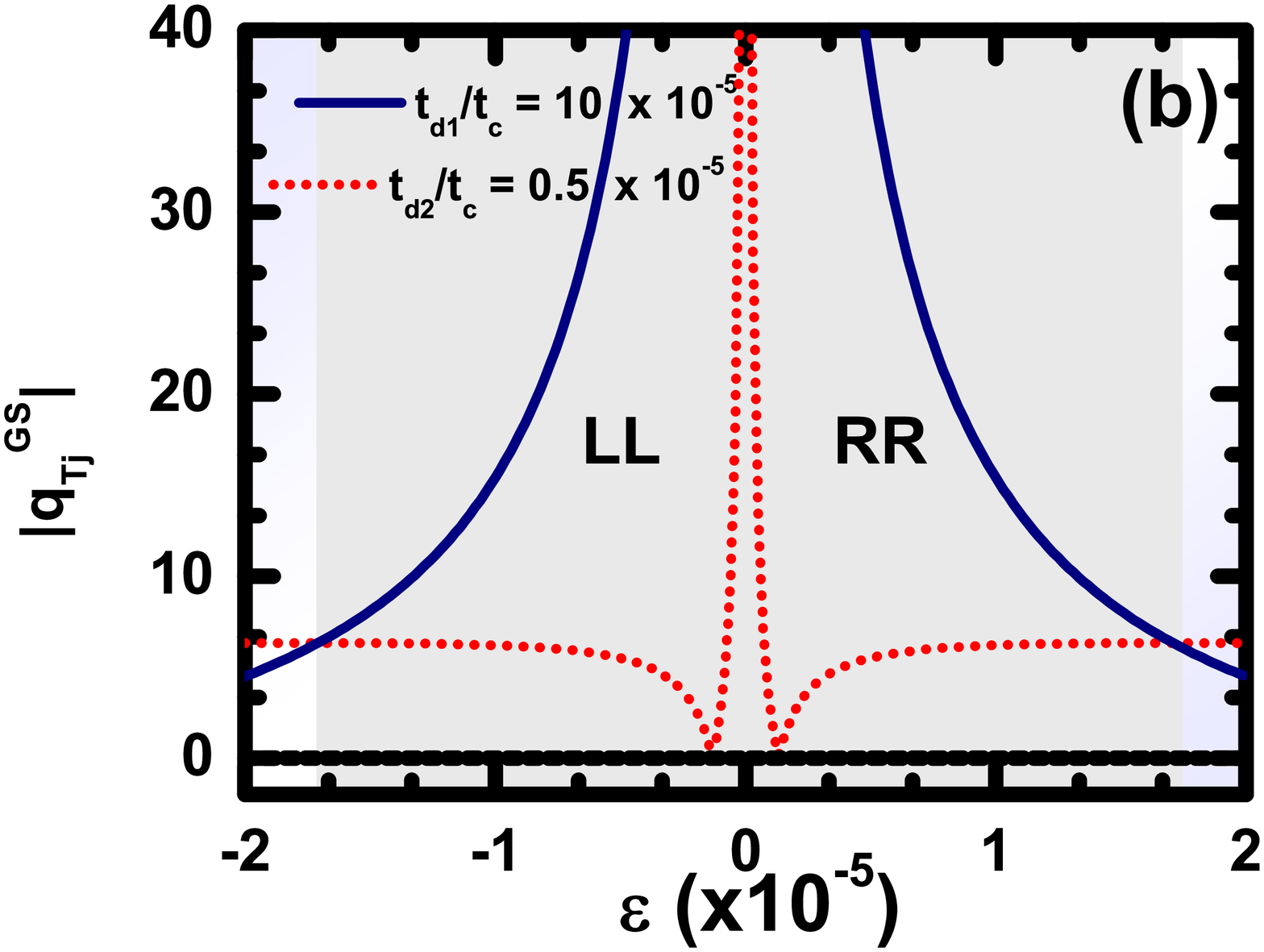}}}
\caption{\label{Fig2TLast}(Color online) Absolute value of the Fano parameter of Eq. (\ref{eq:Fano_jGS})
as a function of the energy $\varepsilon$ for $v_{0}=0.14$ and $t_{d2}/t_{c}=0.5\times10^{-5}$.
(a) $t_{d1}/t_{c}=1\times10^{-5}$. (b) $t_{d1}/t_{c}=10\times10^{-5}$.
In these graphs we have defined a shaded region in which the condition
$|q_{T2}^{GS}|<|q_{T1}^{GS}|$ is verified. By increasing the ratio
$t_{d1}/t_{c}$ from $1\times10^{-5}$ to $10\times10^{-5}$ this
region is enlarged as shown in (b). For both cases it is observed
that the Fano factor tends to infinity as the energy approaches the
Fermi level. }
\end{figure}

In Fig. \ref{Fig2TLast}, we present the absolute value of the Fano
parameter $\left|q_{Tj}^{GS}\right|$ of Eq. (\ref{eq:Fano_jGS})
as a function of the energy $\varepsilon$. All curves exhibit a general
trend in which the Fano factor tends to infinity as the energy approaches
the Fermi level ($\varepsilon=0$) and decays to a finite value as
$\varepsilon$ increases towards the band edge. By increasing the
ratio $t_{dj}/t_{c}$ the Fano parameter diverges more rapidly. In
Fig. \ref{Fig2TLast}(a) the curve for $|q_{T1}^{GS}|$ (solid-blue
curve) becomes broader than the curve for $|q_{T2}^{GS}|$ (dotted-red
curve) for $|\varepsilon|\lesssim0.2$ and keeps increasing as $\varepsilon\rightarrow0$.
This behavior becomes even more pronounced at Fig. \ref{Fig2TLast}(b) where the ratio
$t_{d1}/t_{c}$ is increased 10 times while $t_{d2}/t_{c}$ is kept
fixed. As a result, $|q_{T1}^{GS}|$ lies above the $|q_{T2}^{GS}|$
almost the entire range except in the borders of the scale for $|\varepsilon|\gtrsim1.7$.
The shaded regions in Fig. \ref{Fig2TLast} are defined in such a
way that $|q_{T2}^{GS}|<|q_{T1}^{GS}|$ while out of these regions
the opposite relation is verified, i.e., $|q_{T2}^{GS}|>|q_{T1}^{GS}|$.
In the last case, the resulting behavior yields a Fano interference
in the LDOS dictated by the subsurface impurity, where $|q_{T2}^{GS}|$
becomes dominant. It is worth mentioning that in spite of the condition
$t_{d1}/t_{c}>t_{d2}/t_{c}$ is maintained for all curves in Fig.
\ref{Fig2TLast}, there are regions of $\varepsilon$ in which the
opposite condition ($|q_{T2}^{GS}|>|q_{T1}^{GS}|$) is verified. This
unexpected feature is a result of the interplay between the quantum interference in the double impurity system and peculiar behavior of the graphene density of states. We point out that
such a behavior is not present in the 2DEG setup as we will verify
in Sec. \ref{sub:5b}.

In Fig. \ref{Fig2TLast}(b), we move the STM tip closer to the host
by choosing $t_{d1}/t_{c}=10\times10^{-5}$. Within the shaded regions,
the adatom gives the dominant impact to the interference and $|q_{T1}^{GS}|$ (solid-blue
curve) overcomes $|q_{T2}^{GS}|$ (dotted-red curve) . When $t_{dj}/t_{c}\ll1$,
the profile of the LDOS for the graphene is expected to exhibit resonances.
This result contrasts the 2DEG system, where standard Fano's theory
predicts antiresonances. In fact, Eq. (\ref{eq:Fano_j2DEG}) ensures
$\left|q_{Tj}^{2DEG}\right|\rightarrow0$ for $\varepsilon\rightarrow0$
 \cite{Fano1,Fano2}.

\begin{figure}[!]
\centerline{\resizebox{3.5in}{!}{ \includegraphics[clip,width=0.6\textwidth]{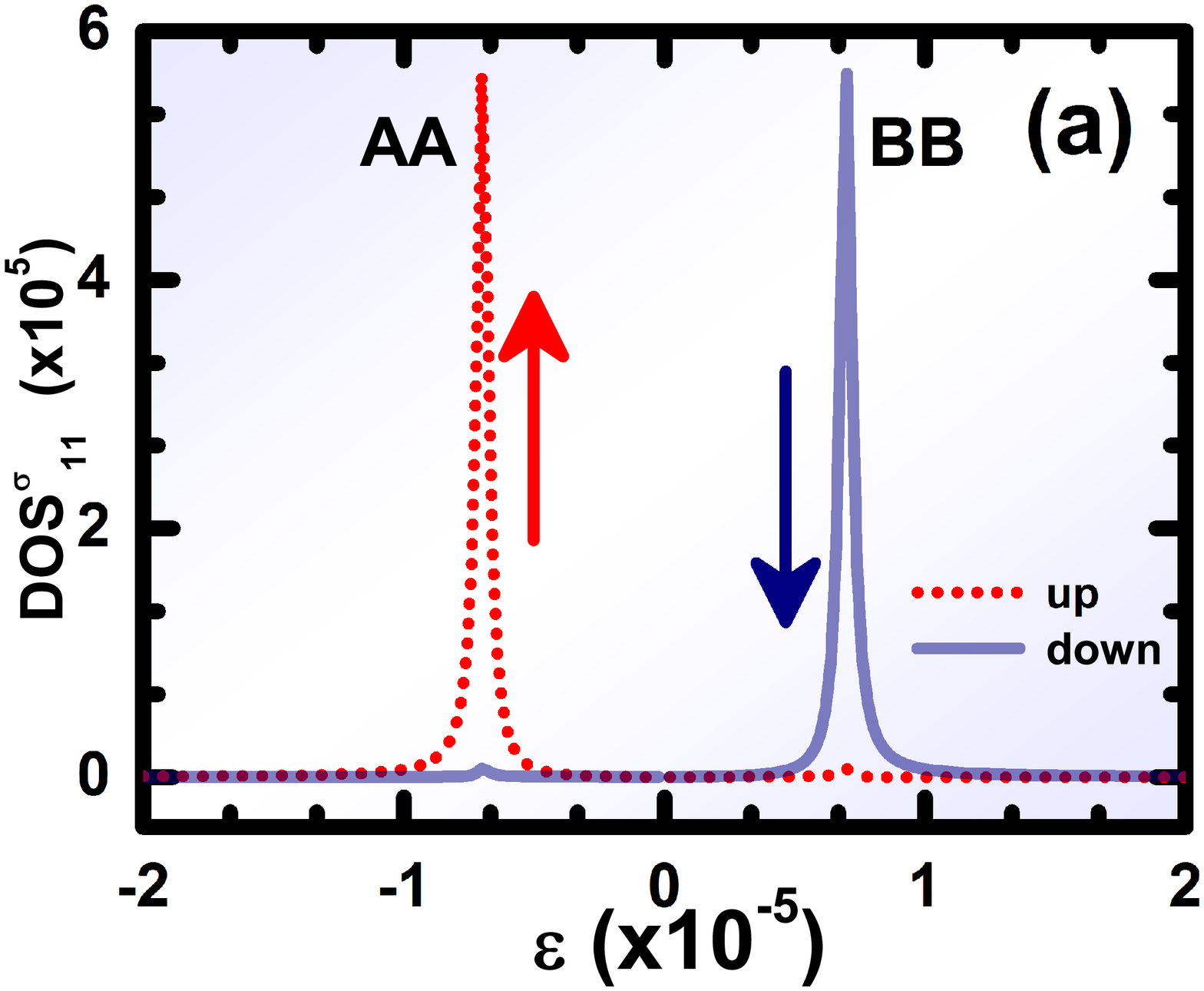}}}
\centerline{\resizebox{3.5in}{!}{ \includegraphics[clip,width=0.6\textwidth]{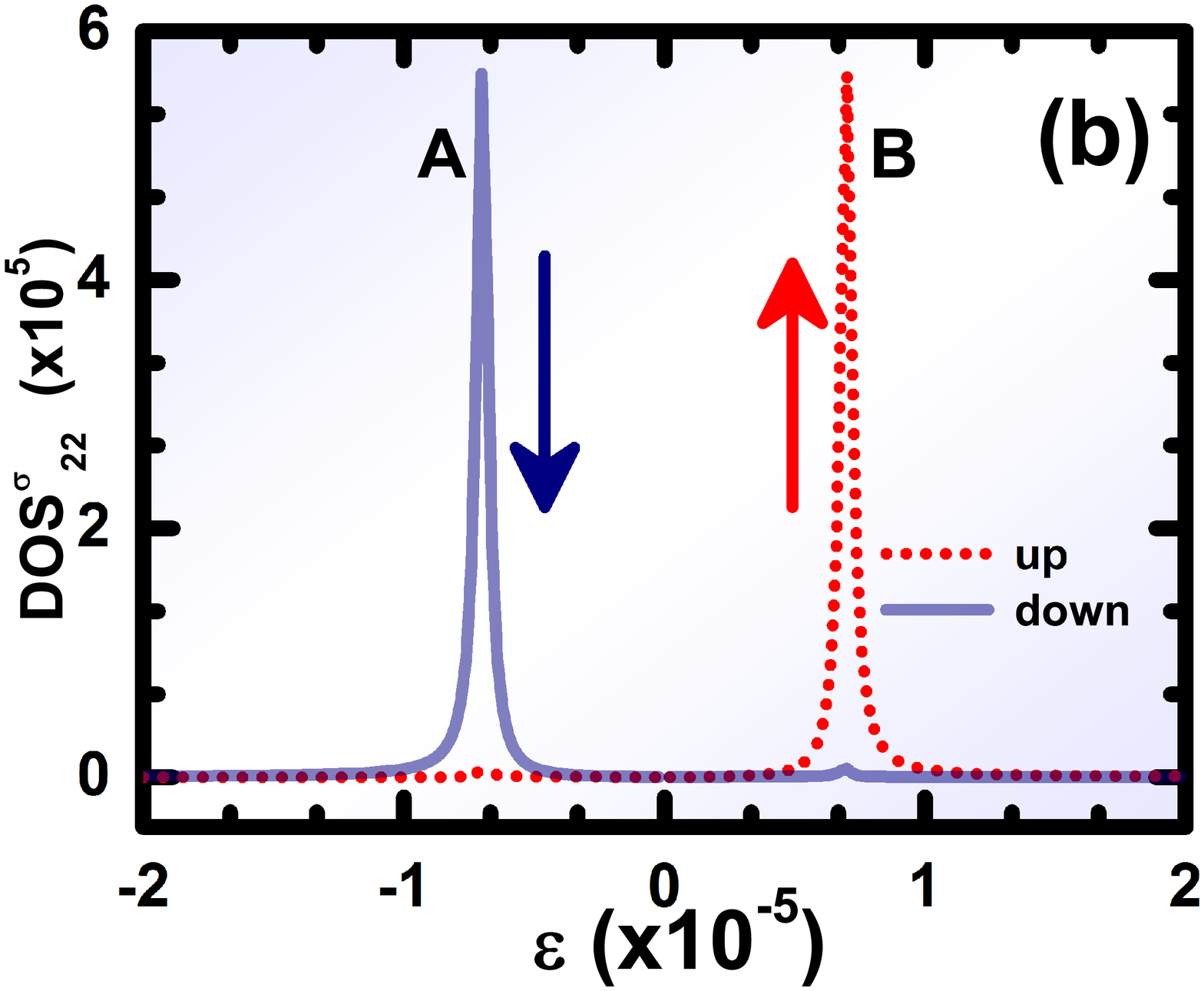}}}
\caption{\label{Fig3TLast}(Color online) Parameters: $v_{0}=0.14$ and $\varepsilon_{1d}=\varepsilon_{2d}=0$.
Here we use Eq. (\ref{eq:DOSjj}) for the DOS of the impurities. (a)
DOS for the adatom with Zeeman energy $\Delta_{1}=2\times10^{-5}$.
(b) DOS for the subsurface impurity with Zeeman energy $\Delta_{2}=-\Delta_{1}$
{[}antiparallel magnetic fields{]}. The arrows (red and blue) indicate
a given spin corresponding to the resonance.}
\end{figure}

As it was mentioned earlier, the spin components of the
DOSs for the impurities {[}Eq. (\ref{eq:DOSjj}){]} are displaced
in opposite directions away from the Fermi level ($\varepsilon=0$)
as Eq. (\ref{eq:E2_up_down}) ensures. In Fig. \ref{Fig3TLast}(a),
the Zeeman energy is $\Delta_{1}=2\times10^{-5}$, thus the resonance
of the localized state in the adatom for spin-up (dotted-red curve)
moves to the left, while the corresponding for spin-down (solid-blue
curve) goes to the right {[}see the ``down'' and ``up'' drawn
arrows in this figure{]}. We identify such resonances by the letters
``AA'' and ``BB''. For the subsurface impurity, $\Delta_{2}=-\Delta_{1}$
and the displacements of the peaks become reversed as displayed in
Fig. \ref{Fig3TLast}(b) {[}the ``down'' and ``up'' drawn arrows
illustrate such a process{]}. These peaks are labeled by the names
``A'' and ``B''. The Fano factors shown in Fig. \ref{Fig2TLast}
and the peaks ``A'', ``B'', ``AA'' and ``BB'' for the spin-dependent
resonances will help us to perceive, in the asymmetric limit $t_{d1}/t_{c}\neq t_{d2}/t_{c}$,
the reversal of the majority spin-component in the LDOS.

\begin{figure}[h]
\centerline{\resizebox{3.5in}{!}{ \includegraphics[clip,width=0.6\textwidth]{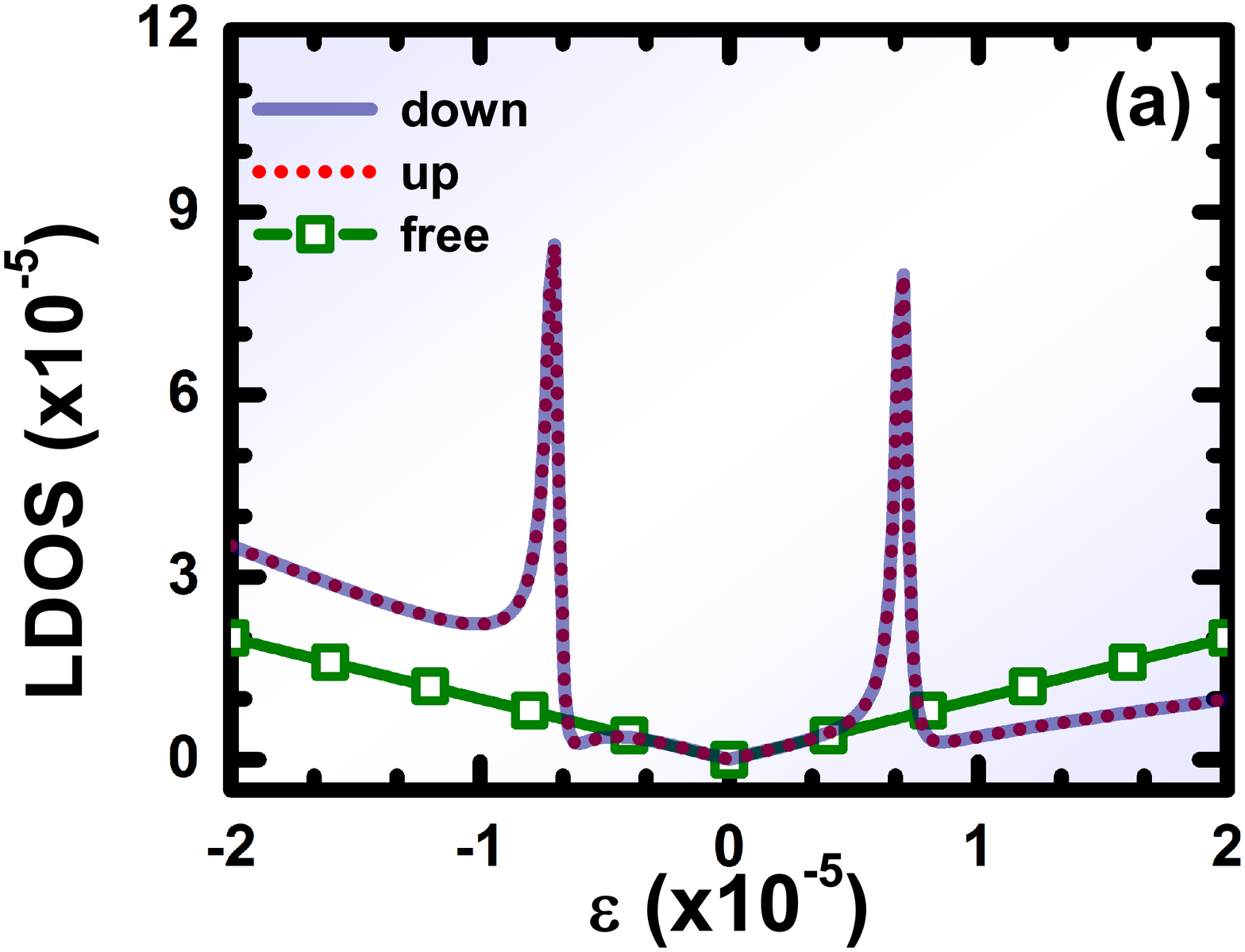}}}
\centerline{\resizebox{3.5in}{!}{ \includegraphics[clip,width=0.6\textwidth]{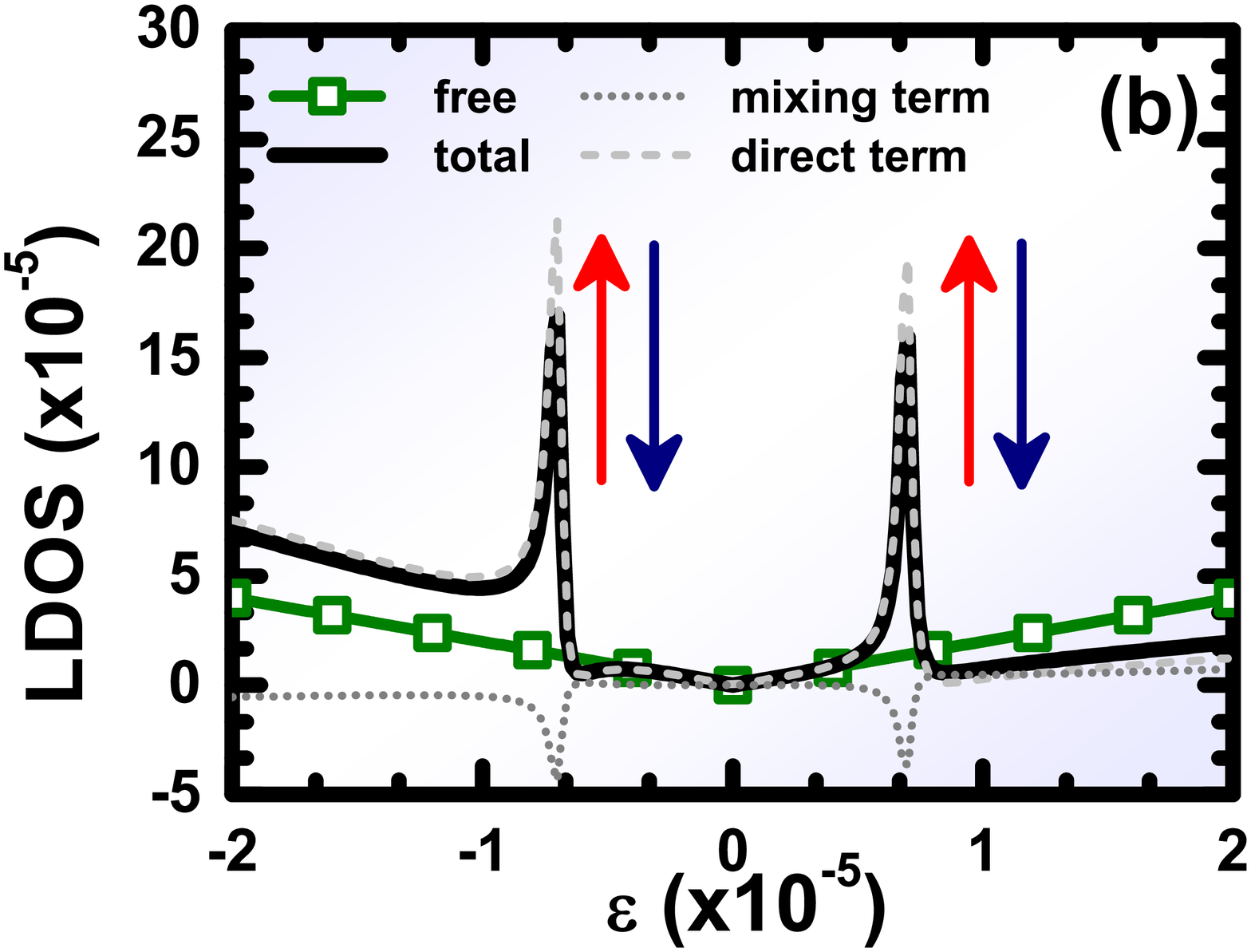}}}
\caption{\label{Fig4TLast}(Color online) Parameters: $v_{0}=0.14$, $\varepsilon_{1d}=\varepsilon_{2d}=0$,
$t_{d1}/t_{c}=t_{d2}/t_{c}=1\times10^{-5}$ {[}symmetric limit of
Fano factors{]} and $\Delta_{1}=-\Delta_{2}=2\times10^{-5}$ {[}antiparallel
magnetic fields{]}. (a) In the symmetric limit, $\rho_{LDOS}^{\sigma}$
{[}Eq. (\ref{eq:LDOS_split}){]} is spin-degenerate. As a result,
the dotted-red curve and solid-blue curve for spins up and down, respectively,
are superimposed. For comparison, the DOS for graphene free of impurities
{[}Eq. (\ref{eq:LDOS_free}){]} is represented by the green line with
squares. (b) The solid-black curve represents the total LDOS given
by sum of spin-up and down contributions {[}Eq. (\ref{eq:LDOS_sd}){]}.
Arrows are included in order to illustrate the spins corresponding
to the resonance. The green curve with squares is the same as Fig.
\ref{Fig4TLast}(a). Additionally, the dashed-gray and dotted-gray
lines are the plots of Eqs. (\ref{eq:LDOS_p1}) and (\ref{eq:LDOS_p2}),
respectively. These equations correspond to the direct and mixed contributions
for the total LDOS under the presence of the impurities. }
\end{figure}

\begin{figure}[h]
\centerline{\resizebox{3.5in}{!}{ \includegraphics[clip,width=0.6\textwidth]{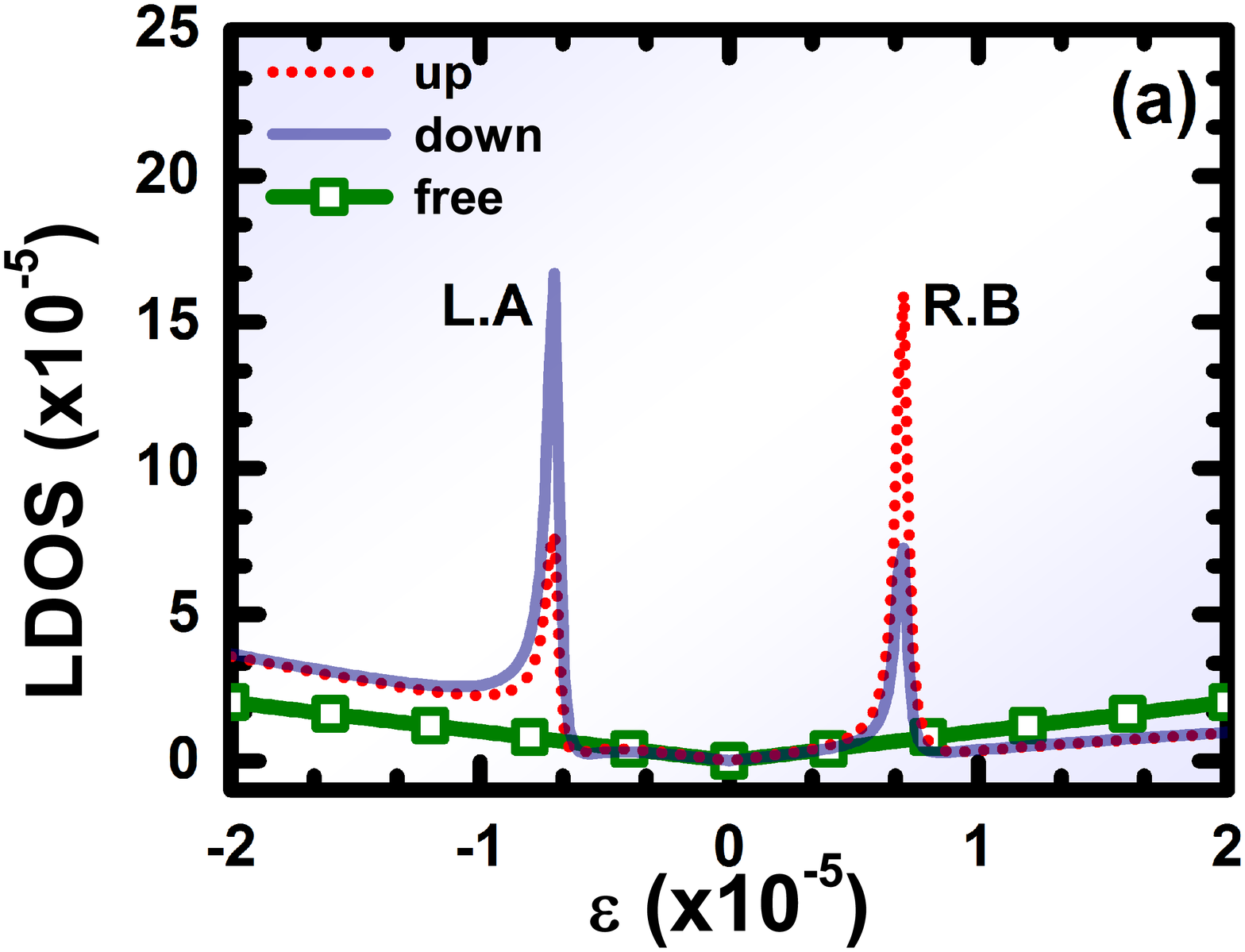}}}
\centerline{\resizebox{3.5in}{!}{ \includegraphics[clip,width=0.6\textwidth]{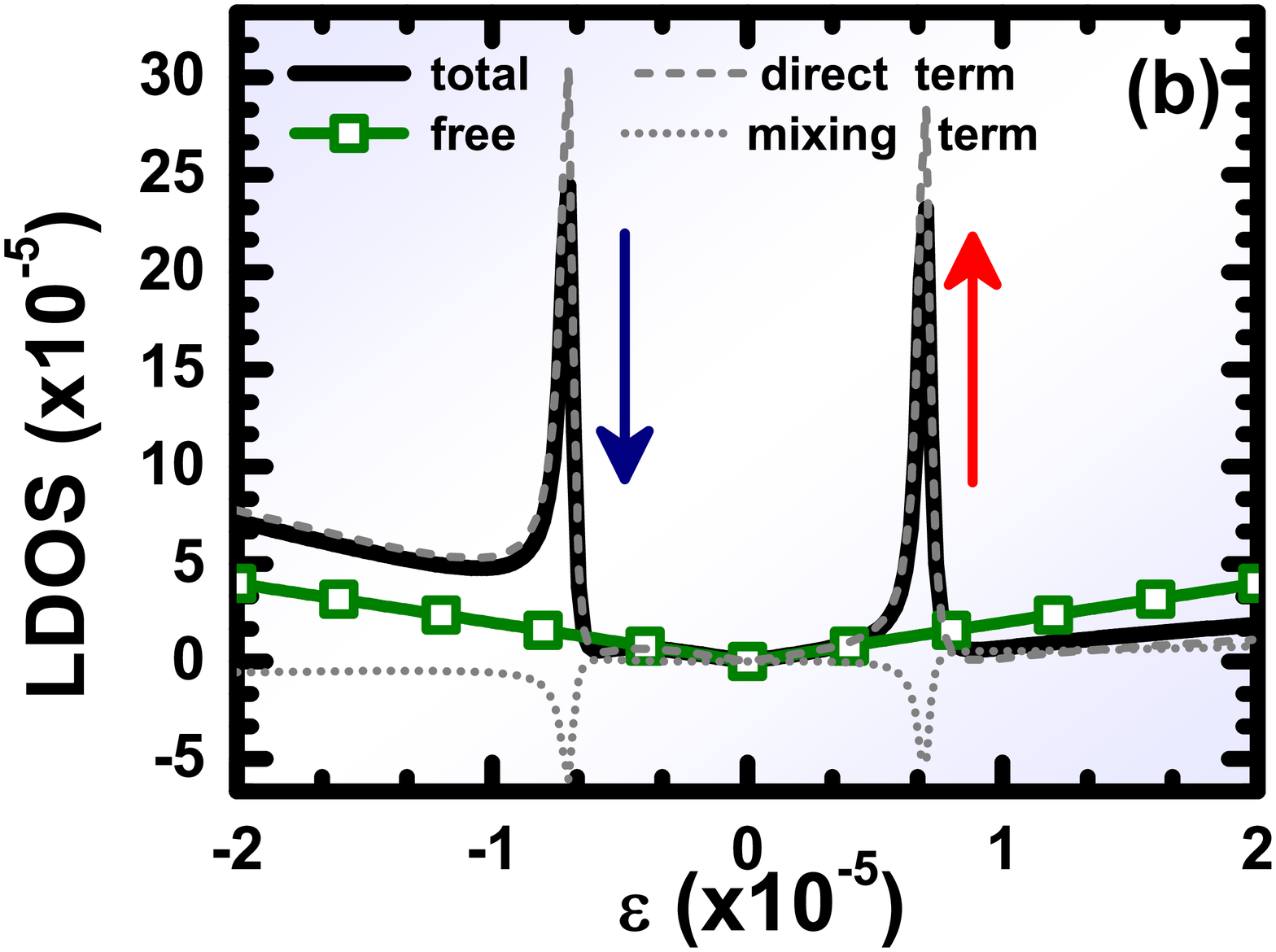}}}
\caption{\label{Fig5TLast}(Color online) Parameters: $v_{0}=0.14$, $\varepsilon_{1d}=\varepsilon_{2d}=0$,
$t_{d1}/t_{c}=1\times10^{-5}$ and $t_{d2}/t_{c}=0.5\times10^{-5}$
{[}asymmetric limit of Fano factors{]}, with $\Delta_{1}=-\Delta_{2}=2\times10^{-5}$
{[}antiparallel magnetic fields{]}. (a) In the asymmetric limit, $\rho_{LDOS}^{\sigma}$
{[}Eq. (\ref{eq:LDOS_split}){]} becomes spin-dependent which is evident
by the peaks with different amplitudes for spin-up (dotted-red curve)
and spin-down (solid-blue curve). For comparison, the DOS for graphene
free of impurities {[}Eq. (\ref{eq:LDOS_free}){]} is represented
by the green line with squares. (b) By summing the curves of Fig.
\ref{Fig5TLast}(a) for spin-up and down one obtains the solid-black
curve for the total LDOS {[}Eq. (\ref{eq:LDOS_sd}){]}. In contrast
to the results of Fig. \ref{Fig4TLast}(b), now the total LDOS exhibits
a spin polarization as indicated by the arrows at each peak. It is
also shown the dashed-gray and dotted-gray lines for the plots of
Eqs. (\ref{eq:LDOS_p1}) and (\ref{eq:LDOS_p2}), respectively. Notice
that in this case the direct contribution is stronger than the symmetric
case of Fig. \ref{Fig4TLast}.}
\end{figure}

\begin{figure}[h]
\centerline{\resizebox{3.5in}{!}{ \includegraphics[clip,width=0.6\textwidth]{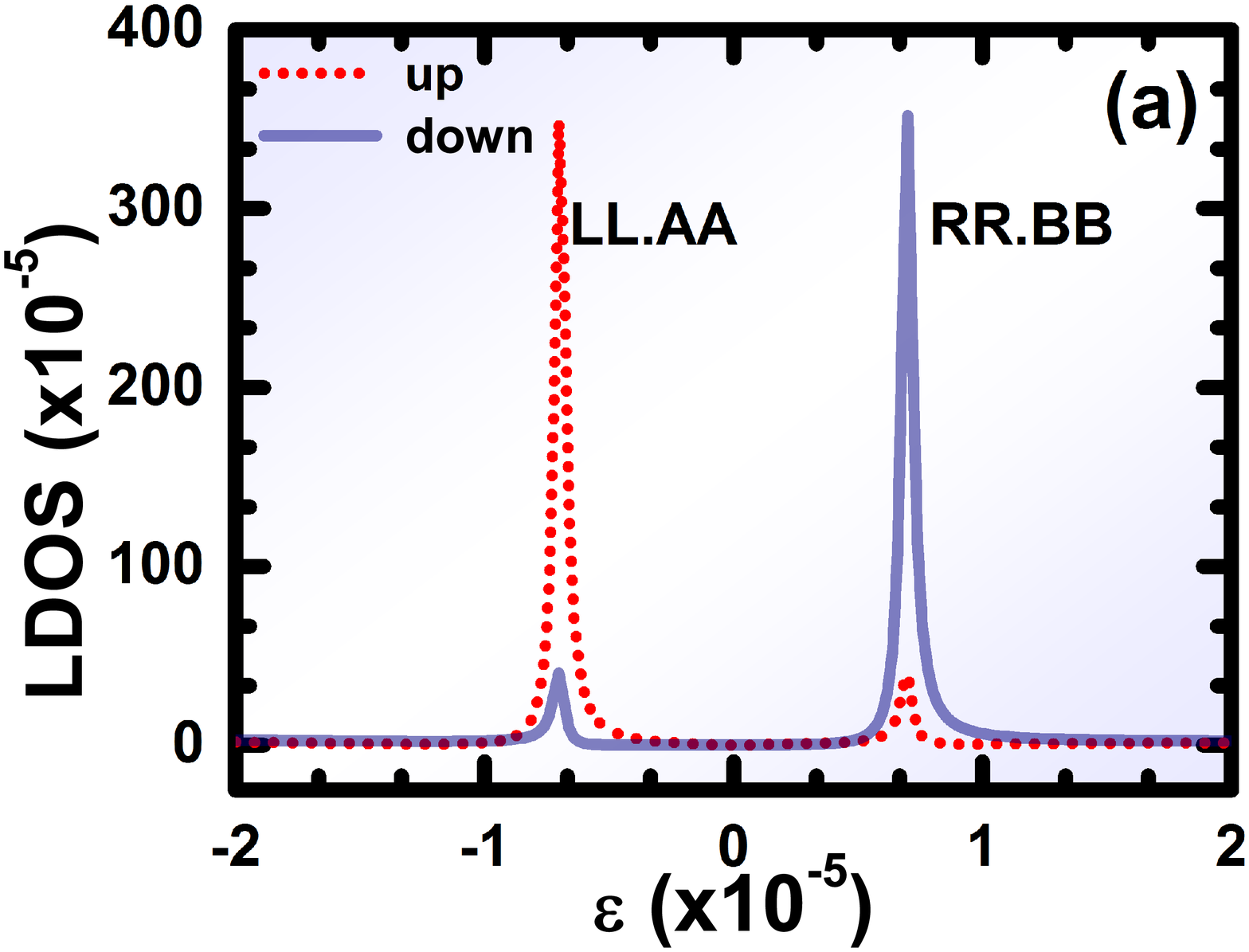}}}
\centerline{\resizebox{3.5in}{!}{ \includegraphics[clip,width=0.6\textwidth]{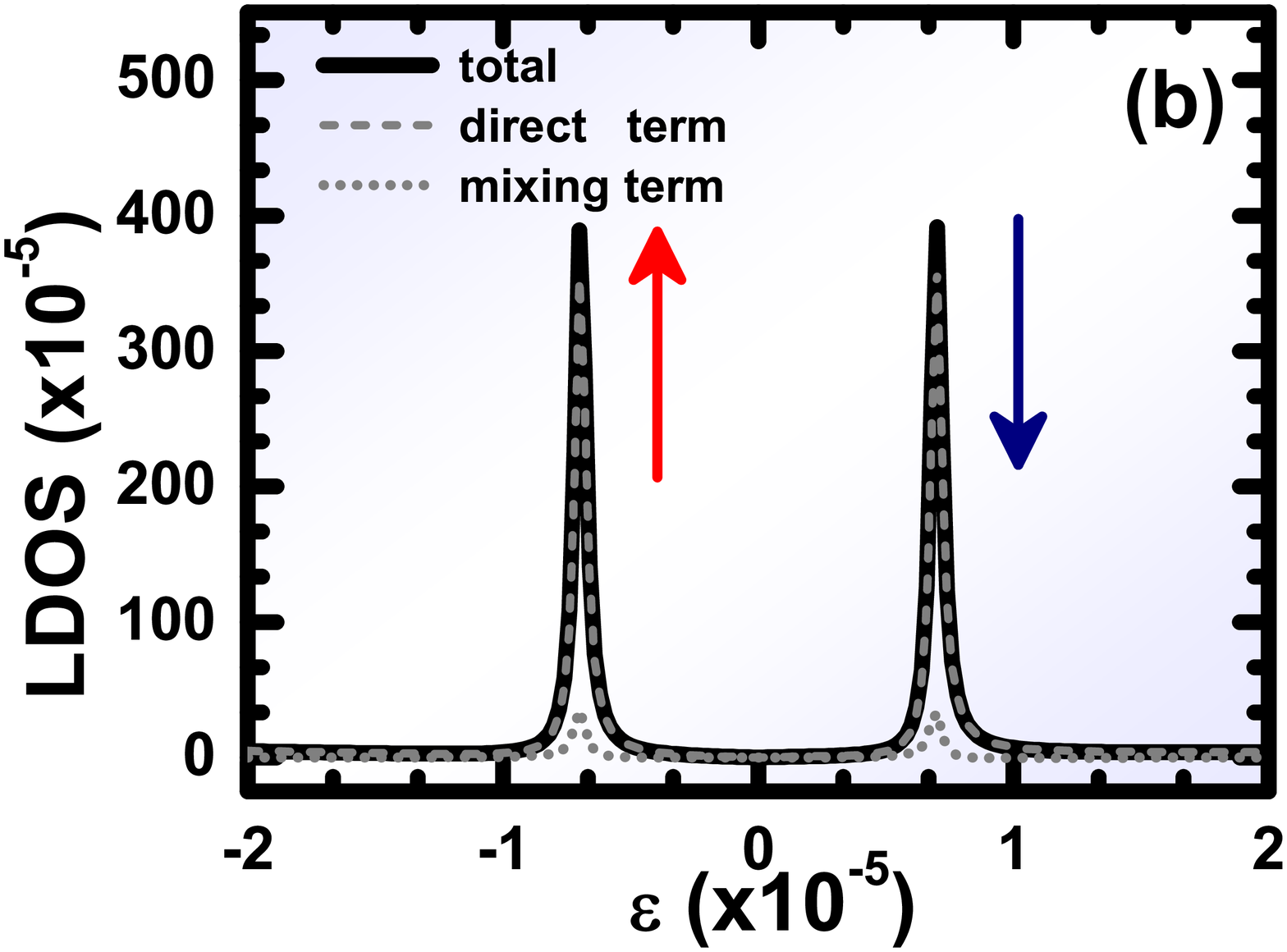}}}
\caption{\label{Fig6TLast}(Color online) Parameters: $v_{0}=0.14$, $\varepsilon_{1d}=\varepsilon_{2d}=0$,
$t_{d1}/t_{c}=10\times10^{-5}$ and $t_{d2}/t_{c}=0.5\times10^{-5}$
{[}asymmetric limit of Fano factors{]}, with $\Delta_{1}=-\Delta_{2}=2\times10^{-5}$
{[}antiparallel magnetic fields{]}. (a) In the asymmetric limit, $\rho_{LDOS}^{\sigma}$
{[}Eq. (\ref{eq:LDOS_split}){]} is spin-dependent as one can observe
by comparing the curves for spin-up (red-dotted curve) and spin-down
(solid-blue curve). (b) The total density of states LDOS of Eq. (\ref{eq:LDOS_sd})
(solid-black curve) and the contributions of the direct (dashed-gray
curve) and mixing (dotted-gray curve) LDOS obtained from Eqs. (\ref{eq:LDOS_p1})
and (\ref{eq:LDOS_p2}). The arrows indicate the majority spin corresponding
to the resonance. The increase of $t_{d1}/t_{c}=10\times10^{-5}$
leads to the inversion of the mixing curve which exhibits resonances
instead of antiresonances as previously shown in Fig. \ref{Fig5TLast}(b).
As a consequence, the majority spin at each peak is also inverted
in comparison to Fig. \ref{Fig5TLast}(b). This shows that is possible
to filter either spin-up or down just varying the distance between
the STM tip and the adatom. }
\end{figure}

\begin{figure}
\centering \includegraphics[width=0.5\textwidth]{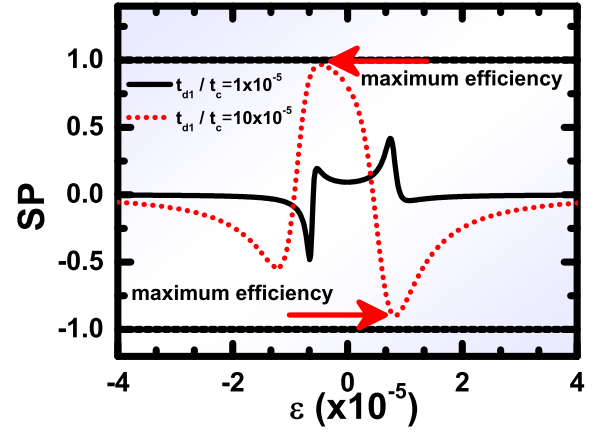}
\caption{\label{Fig7TLast}(Color online) Parameters: $v_{0}=0.14$, $\varepsilon_{1d}=\varepsilon_{2d}=0$,
$t_{d2}/t_{c}=0.5\times10^{-5}$ and $\Delta_{1}=-\Delta_{2}=2\times10^{-5}$
{[}antiparallel magnetic fields{]}. Transport polarization of Eq.
(\ref{eq:SP}) as a function of the energy $\varepsilon$. $t_{d1}/t_{c}=1\times10^{-5}$
for the solid-black curve. $t_{d1}/t_{c}=10\times10^{-5}$ for the
dotted-red curve. By increasing the value of $t_{d1}/t_{c}$ to $10\times10^{-5}$
it is possible to obtain two points at which the efficiency reaches
a maximum value. In particular, for $\varepsilon\simeq-0.5\times10^{-5}$
the efficiency is 100\% which leads to a full polarized current through
the STM system. }
\end{figure}

\begin{figure}[h]
\includegraphics[width=0.5\textwidth]{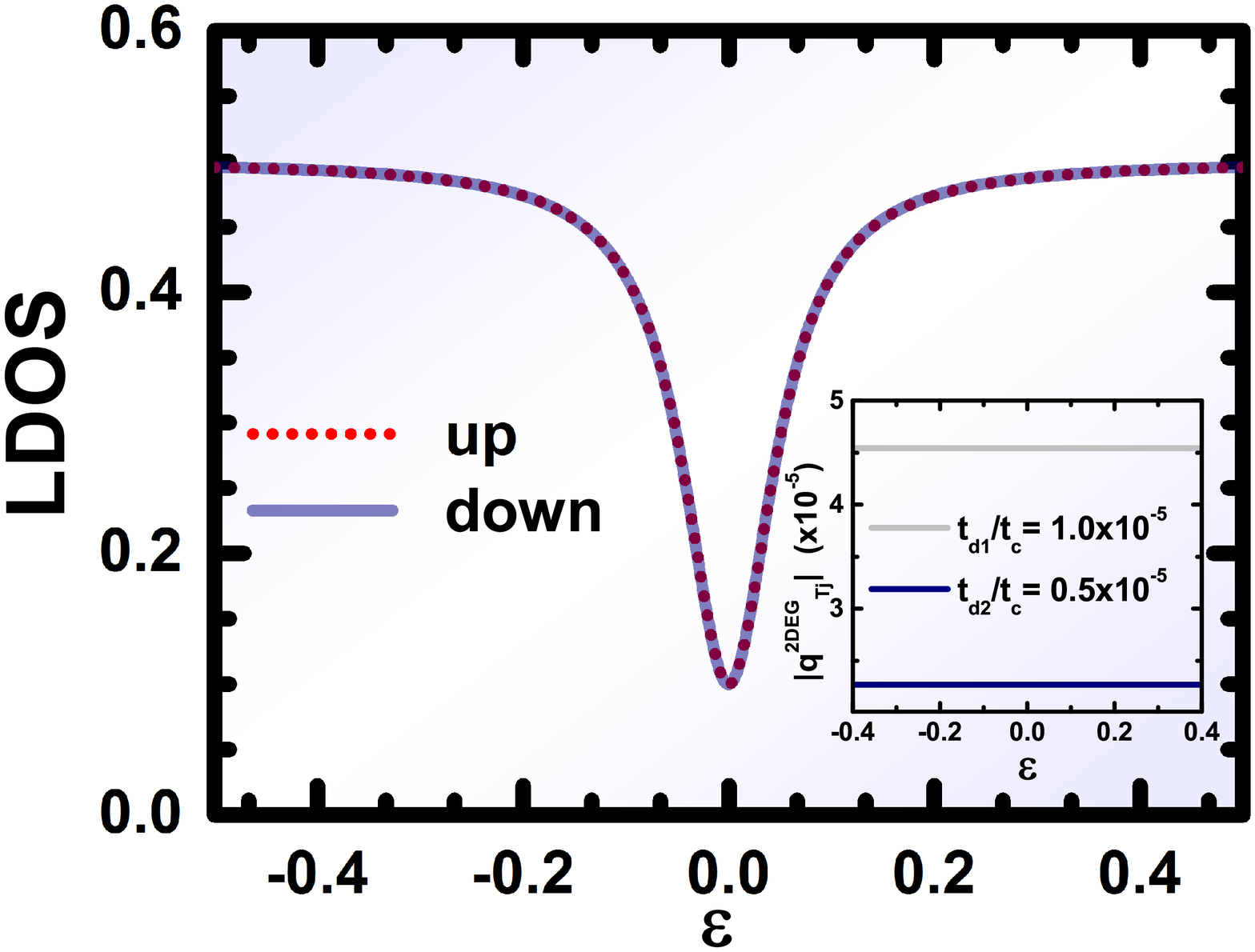} \caption{\label{Fig8TLast}(Color online) Parameters: $v_{0}=0.14$, $\varepsilon_{1d}=\varepsilon_{2d}=0$,
$t_{d1}/t_{c}=1\times10^{-5}$, $t_{d2}/t_{c}=0.5\times10^{-5}$ and
$\Delta_{1}=-\Delta_{2}=2\times10^{-5}$ {[}antiparallel magnetic
fields{]}. It is presented the $\rho_{LDOS}^{\sigma}$ {[}Eq. (\ref{eq:LDOS_split}){]}
as a function of the energy $\varepsilon$ for the 2DEG. Since the
curves for spin-up (dotted-red curve) and spin-down (solid-blue curve)
are superimposed, the total LDOS does not depend on spin as observed
for the graphene sheet. Hence, the 2DEG system does not operate as
a Fano spin-filter. Unlike the graphene, in the 2DEG we always have
$\left|q_{T1}^{2DEG}\right|>\left|q_{T2}^{2DEG}\right|$ for $t_{d1}/t_{c}>t_{d2}/t_{c}$
(inset).}
\end{figure}

As we have antiparallel magnetic fields settled by the constraint
$\Delta_{1}=-\Delta_{2}=2\times10^{-5}$ for the Zeeman energies,
the LDOS does not exhibit spin-dependence in the case of the symmetric
limit of Fano factors, determined by the ratios $t_{d1}/t_{c}=t_{d2}/t_{c}$.
Thus the profile of the LDOS is spin-degenerate as we can see
in Fig. \ref{Fig4TLast}(a) for $t_{d1}/t_{c}=t_{d2}/t_{c}=1\times10^{-5}$,
in which the curves for both spins are superimposed. Despite of the
unpolarized profile of the LDOS, we highlight that the pair of resonances
found is a direct result of the Zeeman splittings at the sites of
the impurities. Otherwise, for levels $\varepsilon_{1d\sigma}=\varepsilon_{2d\sigma}=0$,
the host is completely decoupled from the impurities and the surface
LDOS reduces to Eq. (\ref{eq:LDOS_free}) whose form is illustrated
by the green line with squares also shown in Fig. \ref{Fig4TLast}(a).

In Fig. \ref{Fig4TLast}(b) it is shown the total LDOS (solid-black
curve) obtained from Eq. (\ref{eq:LDOS_sd}). It can be noted that
both spins states contribute equally to the total LDOS for all values
of energy $\varepsilon$. This is represented by two anti-parallel
arrows at both peaks for the total LDOS. Here we also make explicit
the effect of the mixing term $\Sigma_{\sigma}\rho_{1221}^{\sigma}$
obtained from Eq. (\ref{eq:LDOS_p2}) upon the LDOS determined by
Eq. (\ref{eq:LDOS_split}). Fig. \ref{Fig4TLast}(b) shows that the
quantity $\Sigma_{\sigma}\rho_{1221}^{\sigma}$ (dotted-gray curve)
suppresses the direct term $\Sigma_{\sigma}\rho_{1122}^{\sigma}$
calculated from Eq. (\ref{eq:LDOS_p1}). The latter exhibits two resonances
(dashed-gray curve), while the former is characterized by two antiresonances.
The peaks in the direct term are the hallmark of constructive interference,
contrasting to the Fano antiresonances found in the mixing term, which
are signatures of destructive interference. As a result, the total
LDOS of Eq. (\ref{eq:LDOS_sd}) is given by the solid-black curve.
The green line with squares gives $\Sigma_{\sigma}\rho^{GS}\left(\varepsilon\right)$
determined by Eq. (\ref{eq:LDOS_free}), which represents the DOS
of the graphene system in the absence of the impurities where the
peaks are absent as expected.

In Fig. \ref{Fig5TLast}, we analyze the asymmetric limit of the Fano
factors established by the condition $t_{d1}/t_{c}\neq t_{d2}/t_{c}$.
For this situation, with $t_{d1}/t_{c}=1\times10^{-5}$ and $t_{d2}/t_{c}=0.5\times10^{-5}$,
we have verified that the total LDOS becomes spin-dependent. Such
a feature appears in Fig. \ref{Fig5TLast}(a), where the distinction
between the up and down components of the LDOS is evident {[}$\rho_{LDOS}^{\uparrow}\neq\rho_{LDOS}^{\downarrow}${]}.
In the range of negative energies, two aligned peaks with different
amplitudes exist, but the corresponding for spin-down (solid-blue
curve) is more pronounced in respect to the spin-up (dotted-red curve),
i.e., $\rho_{LDOS}^{\downarrow}>\rho_{LDOS}^{\uparrow}$ . At positive
energies, this pattern is reversed {[}$\rho_{LDOS}^{\uparrow}>\rho_{LDOS}^{\downarrow}${]}.
Thus, depending on which resonance peak is probed by the STM tip,
placed at $\varepsilon\simeq-0.7\times10^{-5}$ or at $\varepsilon\simeq0.7\times10^{-5}$,
the system filters predominantly spin-down or up, respectively.

The origin of such a filtering lies within the direct term $\Sigma_{\sigma}\rho_{1122}^{\sigma}$
determined by Eq. (\ref{eq:LDOS_p1}), in particular, it arises from
the contribution of $-\left(q_{T2}^{GS}\right)^{2}{\tt Im}(\tilde{\mathcal{R}}_{d_{2}d_{2}}^{\sigma})=\left(q_{T2}^{GS}\right)^{2}\pi\rho^{GS}\left(D\right)DOS_{22}^{\sigma}$
, where we have used Eq. (\ref{eq:DOSjj}). As $\left|q_{T2}^{GS}\right|>\left|q_{T1}^{GS}\right|$
is valid for the ranges out of the shaded regions {[}sides ``L''
and ``R''{]} in Fig. \ref{Fig2TLast}(a), the parameter$\left(q_{T2}^{GS}\right)^{2}$
enhances the resonances ``A'' and ``B'' of Fig. \ref{Fig3TLast}(b),
thus resulting in the peaks ``L.A'' and ``R.B'' in Fig. \ref{Fig5TLast}(a).

In Fig. \ref{Fig5TLast}(b) we have performed the same analysis as
done for Fig. \ref{Fig4TLast}(b). We have verified that in the asymmetric
limit $t_{d1}/t_{c}\neq t_{d2}/t_{c}$, the mixing term $\Sigma_{\sigma}\rho_{1221}^{\sigma}$
calculated from Eq. (\ref{eq:LDOS_p2}) (dotted-gray curve) suppresses
the peaks of the direct term $\Sigma_{\sigma}\rho_{1122}^{\sigma}$
determined by Eq. (\ref{eq:LDOS_p1}) (dashed-gray curve) as well
as in the symmetric regime $t_{d1}/t_{c}=t_{d2}/t_{c}$ of Fig. \ref{Fig4TLast}(b).
This suppression leads to the solid-black curve, which is obtained
from Eq. (\ref{eq:LDOS_sd}). In contrast to the results of Fig. \ref{Fig4TLast}(b),
in Fig. \ref{Fig5TLast}(b) each black peak of the total LDOS exhibits
a finite polarization whose majority spin is indicated by an arrow
at each peak.

In order to investigate the role of the mixing term $\Sigma_{\sigma}\rho_{1221}^{\sigma}$
upon the total LDOS, we have considered in Fig. (\ref{Fig6TLast})
the STM tip closer to the host surface. To accomplish this situation,
we have increased $t_{d1}/t_{c}$ to $10\times10^{-5}$ keeping $t_{d2}/t_{c}$
fixed to $0.5\times10^{-5}$. Hence, the value of $t_{d1}/t_{c}$
is ten times greater than the corresponding value used in Fig. \ref{Fig5TLast}
which makes the mixing term more relevant in this case. Fig. \ref{Fig6TLast}(a)
exhibits enhanced resonances in respect to those found in Fig. \ref{Fig5TLast}(a).
Notice that the scale of the LDOS axis is also enlarged by a factor
of ten, thus the background DOS of the graphene {[}not displayed in
Fig. \ref{Fig6TLast}(a){]} acts as a flat band within this scale.
Moreover, the main difference between Figs. \ref{Fig5TLast}(a) and
\ref{Fig6TLast}(a) is the reversal of the majority spin for the resonances.

In Fig. \ref{Fig6TLast}(a), the peak at $\varepsilon\simeq-0.7\times10^{-5}$
is dominated by spin-up electrons {[}$\rho_{LDOS}^{\uparrow}>\rho_{LDOS}^{\downarrow}$
in dotted-red curve{]}, while the corresponding at $\varepsilon\simeq+0.7\times10^{-5}$
is due to spin-down {[}$\rho_{LDOS}^{\downarrow}>\rho_{LDOS}^{\uparrow}$
in solid-blue curve{]}. For Fig. \ref{Fig5TLast}(a), we have exactly
the opposite. The origin lies within the term $-\left(q_{T1}^{GS}\right)^{2}{\tt Im}(\tilde{\mathcal{R}}_{d_{1}d_{1}}^{\sigma})=\left(q_{T1}^{GS}\right)^{2}\pi\rho^{GS}\left(D\right)DOS_{11}^{\sigma}$.

In the case of Fig. \ref{Fig6TLast}(a), the corresponding Fano parameters
are described by the curves in Fig. \ref{Fig2TLast}(b) in which $\left|q_{T1}^{GS}\right|>\left|q_{T2}^{GS}\right|$
for the whole energy range {[}shaded regions ``LL'' and ``RR''
in Fig. \ref{Fig2TLast}(b){]}. Thus, the peaks ``AA'' and ``BB''
of Fig. \ref{Fig3TLast}(a) are enhanced by $\left(q_{T1}^{GS}\right)^{2}$
and lead to the new resonances ``LL.AA'' and ``RR.BB'' in Fig.
\ref{Fig6TLast}(a).

We point out that not only the spin-filtering effect becomes reversed,
but also the Fano interference arising from $\Sigma_{\sigma}\rho_{1221}^{\sigma}$
in Eq. (\ref{eq:LDOS_p2}) does. Fig. \ref{Fig6TLast}(b) shows that
the mixing term (dotted-gray curve) is formed by a pair of resonances
which is the opposite pattern as compared to that found in Figs. \ref{Fig4TLast}(b)
and \ref{Fig5TLast}(b). It means that the destructive interference
is replaced by a constructive one. As a result, the direct and mixed
term {[}dashed and dotted-gray curves, respectively{]} are now being
summed leading to the total LDOS {[}Eq. (\ref{eq:LDOS_sd}){]} represented
by the solid-black curve. Once again, we have used an arrow at each
peak to denote down or up majority spin. In summary, the lifting of
the spin-degeneracy in the ``host+impurities'' device is not established
by the Zeeman effect $\Delta_{1}=-\Delta_{2}$, but due to the asymmetric
ratios $t_{d1}/t_{c}\neq t_{d2}/t_{c}$. To the best knowledge, our
work is the first to propose the Fano interference as the mechanism
to filter spins in graphene.

The degree of spin polarization for the transport through the considered setup given by Eq. (\ref{eq:SP})
as a function of the energy $\varepsilon$ is displayed in Fig. \ref{Fig7TLast}.
We have analyzed the cases $t_{d1}/t_{c}=1\times10^{-5}$ (solid-black
curve) and $t_{d1}/t_{c}=10\times10^{-5}$ (dotted-red curve). In
the former situation, for energies below and nearby the Fermi level,
values of positive polarizations {[}$\mathcal{G}^{\uparrow}>\mathcal{G}^{\downarrow}${]}
and negative {[}$\mathcal{G}^{\downarrow}>\mathcal{G}^{\uparrow}${]}
occur, while above the Fermi level, polarization remains positive.
However, it never exceeds $\left|0.5\right|$. In the case $t_{d1}/t_{c}=10\times10^{-5}$,
which mimics the STM tip closer to the host, the pattern of the polarization
observed for $t_{d1}/t_{c}=1\times10^{-5}$ is reversed. Moreover,
it reaches the maximum value $+1$ at $\varepsilon\simeq-0.5\times10^{-5}$
and approaches $-1$ for $\varepsilon\simeq+0.8\times10^{-5}$ {[}see
the horizontal arrows{]}.

\subsection{2DEG system}

\label{sub:5b}

In this section we explore the 2DEG system with standard quadratic dispersion. To this end, we employ Eq. (\ref{eq:LDOS_sd})
for the LDOS by taking into account Eqs. (\ref{eq:LDOS_free_2}),
(\ref{eq:pass3-1}), (\ref{eq:pass9-1}), (\ref{eq:s_e_GS-2}), (\ref{eq:Fano_j2DEG}) and (\ref{eq:A_2DEG}).

Fig. \ref{Fig8TLast} reveals that even in the asymmetric limit, for
which $t_{d1}/t_{c}=1\times10^{-5}$ and $t_{d2}/t_{c}=0.5\times10^{-5}$,
there is no resolved spin-dependence in the Fano profile of the LDOS.
Such a feature can be visualized via dotted-red and solid-blue curves,
which are characterized by degenerate antiresonances for spins up
and down, respectively. These resonances are predictable due to Eq.
(\ref{eq:Fano_j2DEG}) for the Fano parameter, which gives $\left|q_{Tj}^{2DEG}\right|\rightarrow0$
for $\varepsilon\rightarrow0$ {[}see the ``gray'' and ``blue''
lineshapes in the inset, respectively for $t_{d1}/t_{c}=1\times10^{-5}$
and $t_{d2}/t_{c}=0.5\times10^{-5}${]}. It is worth noting that for
the 2DEG, unlike the graphene setup, there is no region in the Fano
versus $\varepsilon$ plot where the condition $\left|q_{T2}^{2DEG}\right|>\left|q_{T1}^{2DEG}\right|$
for $t_{d1}/t_{c}>t_{d2}/t_{c}$ is verified {[}compare the inset
with the results in Fig. \ref{Fig2TLast}{]}. As a result, the 2DEG
setup does not operate as a spin-filter.

\section{CONCLUSIONS}

\label{sec6} In this work, we have proposed a relativistic spin-filter
consisting of an STM setup with a graphene hosting two lateral impurities.
The mechanism through which the STM picks up preferentially a definite spin
is based on quantum Fano interference. A particular feature provided
by this system is the possibility to choose which spin to filter by tuning the distance between the STM tip and the adatom. For particular conditions pure spin currents may be generated
by the proposed setup, which makes it attractive for possible spintronics applications. To our knowledge, there is no equivalent proposal in literature involving STM based
on graphene hosts.

It is worth mentioning that the subsurface impurity plays an important
role on the transport properties in spite of the weak coupling to
the STM tip. In fact, in Fig. \ref{Fig2TLast} there are regions in
which the condition $|q_{T2}^{GS}|>|q_{T1}^{GS}|$ is verified even
with $t_{d1}/t_{c}>t_{d2}/t_{c}$. This is a striking result of quantum
interference and illustrates the subtle quantum properties of graphene based structures.
These results contrasts with 2DEG where the Fano factors always follow
the same trend, i.e., $|q_{T1}^{2DEG}|>|q_{T2}^{2DEG}|$ when $t_{d1}/t_{c}>t_{d2}/t_{c}$.

\begin{acknowledgments}
This work has the support of the Brazilian agencies CNPq, CAPES and
PROPG-PROPe/UNESP and FP7 IRSES project SPINMET.
\end{acknowledgments}


\begin{thebibliography}{10}
\bibitem{key-8}K. S. Novoselov, Rev. Mod. Phys. \textbf{83}, 837
(2011).

\bibitem{key-9}S. Das Sarma, Shaffique Adam, E. H. Hwang, and Enrico
Rossi, Rev. Mod. Phys. \textbf{83}, 407 (2011).

\bibitem{key-100}N. M. R. Peres, Rev. Mod. Phys. \textbf{82}, 2673
(2010).

\bibitem{key-111}A. H. Castro Neto, F. Guinea, N. M. R. Peres, K.
S. Novoselov, and A. K. Geim, Rev. Mod. Phys. \textbf{81}, 109 (2009).

\bibitem{key-11}B. Uchoa, V. N. Kotov, N. M. R. Peres, and A. H.
Castro Neto, Phys. Rev. Lett. \textbf{101}, 026805 (2008).

\bibitem{key-12}B. Uchoa, L. Yang, S.-W. Tsai, N. M. R. Peres, and
A. H. Castro Neto, Phys. Rev. Lett. \textbf{103}, 206804 (2009).

\bibitem{key-15}Z. G. Zhu, and J. Berakdar, Phys. Rev. B \textbf{84},
165105 (2011).

\bibitem{Fazzio}M. P. Lima, A. J. R. da Silva, and A. Fazzio, Phys.
Rev. B \textbf{84}, 245411 (2009).

\bibitem{key-444}V. W. Brar, R. Decker, H. M. Solowan, Y. Wang, L.
Maserati, K. T. Chan, H. Lee, Ç. O. Girit, A. Zettl, S. G. Louie,
M. L. Cohen, and M. F. Crommie, Nat. Phys. \textbf{7}, 43 (2011).

\bibitem{key-445}A. Saffarzadeh, and G. Kirczenow, Phys. Rev. B
\textbf{85}, 245429 (2012).

\bibitem{key-446}F. Hiebel, P. Mallet, J. Y. Veuillen, and L. Magaud,
Phys. Rev. B \textbf{86}, 205421 (2012).

\bibitem{key-447}P. S. Cornaglia, G. Usaj, and C. A. Balseiro, Phys.
Rev. Lett. \textbf{102}, 046801 (2009).

\bibitem{Eelbo1} T. Eelbo, M. Wasniowska, M. Gyamfi, S. Forti, U. Starke, and R. Wiesendanger
Phys. Rev. B \textbf{87}, 205443 (2013)

\bibitem{Eelbo2} T. Eelbo, M. Wasniowska, P. Thakur, M. Gyamfi, B. Sachs, T. O. Wehling, S. Forti, U. Starke, C. Tieg, A. I. Lichtenstein, and R. Wiesendanger, Phys. Rev. Lett. \textbf{110}, 136804 (2013)

\bibitem{Hardcastle} T. P. Hardcastle, C. R. Seabourne, R. Zan, R. M. D. Brydson, U. Bangert, Q. M. Ramasse, K. S. Novoselov, and A. J. Scott, Phys. Rev. B \textbf{87}, 195430 (2013)

\bibitem{Virgus} Y. Virgus, W. Purwanto, H. Krakauer, and S. Zhang
Phys. Rev. B \textbf{86}, 241406 (2013)

\bibitem{Rudenko} A. N. Rudenko, F. J. Keil, M. I. Katsnelson, and A. I. Lichtenstein
Phys. Rev. B \textbf{86}, 075422 (2012)

\bibitem{Uchoa} B. Uchoa, V. N. Kotov, N. M. R. Peres, and A. H. Castro Neto,
Phys. Rev. Lett. \textbf{101}, 026805 (2008)

\bibitem{Chan} K.T. Chan, J.B. Neaton, and M.L. Cohen
Phys. Rev. B \textbf{77}, 235430 (2008)

\bibitem{Huang} Bing Huang, Jaejun Yu, and Su-Huai Wei
Phys. Rev. B 84, 075415 (2011)

\bibitem{Hong} X. Hong, S.-H. Cheng, C. Herding, and J. Zhu
Phys. Rev. B 83, 085410 (2011)

\bibitem{Lehtinen} P. O. Lehtinen, A. S. Foster, A. Ayuela, A. Krasheninnikov, K. Nordlund, and R. M. Nieminen
Phys. Rev. Lett. 91, 017202 (2003)

\bibitem{Chen} J.H. Chen, C. Jang, S. Adam, M.S. Fuhrer, E.D. Williams, and
M. Ishigami, Nature Phys. \textbf{4}, 377 (2008)

\bibitem{Pi} K. Pi, K. M. McCreary, W. Bao, W. Han, Y. F. Chiang, Y. Li, S.-W.
Tsai, C. N. Lau, and R. K. Kawakami, Phys.Rev.B \textbf{80}, 075406 (2009)

\bibitem {Alemani} M. Alemani, A. Barfuss, B. Geng, C. Girit, P. Reisenauer, M. F. Crommie, F. Wang, A. Zettl, and F. Hellman, Phys. Rev. B \textbf{86}, 075433 (2012)

\bibitem{Schedin} F. Schedin, A. Geim, S. Morozov, E. Hill, P. Blake, M. Katsnelson, and K. Novoselov, Nature Mater. \textbf{6}, 652 (2007)

\bibitem{Wehling}  T.O. Wehling, K.S. Novoselov, S.V. Morozov, E.E. Vdovin, M.I. Katsnelson, A. K. Geim, and A. I. Lichtenstein, Nano Lett. \textbf{8}, 173 (2008).

\bibitem{Gyamfi1} M. Gyamfi, T. Eelbo, M. Wasniowska, T. O. Wehling, S. Forti, U. Starke, A. I. Lichtenstein, M.I. Katsnelson, and R. Wiesendanger, Phys. Rev. B \textbf{85}, 161406 (2012)

\bibitem{Gyamfi2} M. Gyamfi, T. Eelbo, M. Was'niowska, and R. Wiesendanger
Phys. Rev. B 84, 113403 (2011)

\bibitem{Brar} V.W. Brar, R. Decker, H. Solowan, Y. Wang, L. Maserati, K.T.
Chan, H. Lee, C.O. Girit, A. Zettl, S. G. Louie, M.L. Cohen, and M. F. Crommie,
Nat. Phys. \textbf{7}, 43 (2011)

\bibitem{key-1}J. Tersoff, and D. R. Hamann, Phys. Rev. Lett. \textbf{50},
1998 (1983).

\bibitem{key-2}J. Tersoff, and D. R. Hamann, Phys. Rev. B \textbf{31},
805 1985.

\bibitem{Fano1} U. Fano, Phys. Rev. \textbf{124}, 1866 (1961).

\bibitem{Fano2}A. E. Miroshnichenko, S. Flach, and Y. S. Kivshar,
Rev. Mod. Phys. \textbf{82},2257 (2010).

\bibitem{Hewson} A. C. Hewson, The Kondo Problem to Heavy Fermions,
(Cambridge University Press, Cambridge, England 1993).

\bibitem{STM4} H. C. Manoharan, C. P. Lutz, and D. M. Eigler, Nature
\textbf{403}, 512 (2000).

\bibitem{STM5} V. Madhavan, W. Chen, T. Jamneala, and F. Crommie,
Phys. Rev. B \textbf{64}, 165412 (2001).

\bibitem{STM6} N. Knorr, M. A. Schneider, L. Diekhöner, P. Wahl and
K. Kern, Phys. Rev. Lett. \textbf{88}, 096804 (2002).

\bibitem{STM12} A. F. Otte, M. Ternes, K. V. Bergmann, S. Loth, H.
Brune, C. P. Lutz, C. F. Hirjibehedin, and A. J. Heinrich, Nature
Physics \textbf{4}, 847 (2008).

\bibitem{STM13} M. Ternes, A. J. Heinrich and W. D. Schneider, J.
Phys.: Condens. Matter \textbf{21}, 053001 (2009).

\bibitem{Seridonio}A. C. Seridonio, F. S. Orahcio, F. M. Souza, and
M. S. Figueira, Phys. Rev. B \textbf{85}, 165109 (2012).

\bibitem{Kawahara}S. L. Kawahara, J. Lagoute, C. Chacon, Y. Girard,
J. Klein and S. Rousset, Phys. Rev. B \textbf{82} 020406R (2010).

\bibitem{SPSTM1}A. C. Seridonio, F. M. Souza, and I. A. Shelykh,
J. Phys.: Condens. Matter \textbf{21}, 095003 (2009).

\bibitem{SPSTM2}A. C. Seridonio, F. M. Souza, J. Del Nero, and I.
A. Shelykh, Physica E \textbf{41}, 1611 (2009).

\bibitem{Poliana} P. H. Penteado, F. M. Souza, A. C. Seridonio, E.
Vernek, and J. C. Egues, Phys. Rev. B \textbf{84}, 125439, (2011).

\bibitem{SPSTM5} N. Néel, J. Kröger, and R. Berndt, Phys. Rev. B
\textbf{82}, 233401 (2010).

\bibitem{Yunong} Y. Qi, J. X. Zhu, S. Zhang, and C. S. Ting, Phys.
Rev. B \textbf{78} 045305 (2008).

\bibitem{FM88} M. Sindel, L. Borda,J. Martinek, R. Bulla, J. König,
G. Schön, S. Maekawa, and J. von Delft, Phys. Rev. B \textbf{76},
045321 (2007).

\bibitem{FM9} K. Hamaya, M. Kitabatake, K. Shibata, M. Jung, M. Kawamura,
K. Hirakawa, T. Machida, and T. Taniyama, Appl. Phys. Lett. \textbf{91},
232105 (2007).

\bibitem{FM10} K. Hamaya, M. Kitabatake, K. Shibata, M. Jung, M.
Kawamura, S. Ishida, T. Taniyama, K. Hirakawa, Y. Arakawa, and T.
Machida, Phys. Rev. B \textbf{77}, 081302(R) (2008).

\bibitem{FM11} M. R. Calvo, J. F. Rossier, J.J. Palacios, D. Jacob,
D. Natelson, and C. Untiedt, Nature \textbf{458}, 1150 (2009).

\bibitem{FM12} J. Hauptmann, J. Paaske, and P. Lindelof, Nature Phys.
\textbf{4}, 373 (2008).

\bibitem{new1} I. Weymann, and L. Borda, Phys. Rev. B \textbf{81},
115445 (2010).

\bibitem{new2} I. Weymann, Phys. Rev. B \textbf{83}, 113306 (2011).

\bibitem{new3} M. Misiorny, I. Weymann and J. Barna{\'{s}}, Phys.
Rev. Lett. \textbf{106}, 126602 (2011).

\bibitem{new4} M. Gaass, A. K. Hütel, K. Kang, I. Weymann, J. von
Delft, and Ch. Strunk, Phys. Rev. Lett. \textbf{107}, 176808 (2011).

\bibitem{new5} M. Misiorny, I. Weymann, and J. Barna{\'{s}}, Phys.
Rev. B \textbf{84}, 035445 (2011).

\bibitem{key-10}M. E. Torio, K. Hallberg, S. Flach, A. E. Miroshnichenko,
and M. Titov, Eur. Phys. J. B \textbf{37}, 399 (2004).

\bibitem{SIAM} P. W. Anderson, Phys. Rev. \textbf{124}, 41 (1961).

\bibitem{QD1}D. Goldhaber-Gordon, H. Shtrikman, D. Mahalu, D. Abusch-
Magder, U. Meirav, and M. A. Kastner, Nature \textbf{391}, 156 (1998).

\bibitem{QD2}S. M. Cronenwett, T. H. Oosterkamp, and L. P. Kouwenhoven,
Science \textbf{281}, 540 (1998).

\bibitem{book2} H. Haug and A. P. Jauho, Quantum Kinetics in Transport
and Optics of Semiconductors, Springer series in Solid-State Sciences
123 (Springer, New York, 1996).

\end{thebibliography}
\end{document}